\documentclass[11pt,a4paper]{article}
\pdfoutput=1 

\usepackage{jcappub} 

\usepackage[T1]{fontenc} 
\usepackage{graphicx,subfig,epsfig,epsf,color, bm}
\usepackage{amssymb,amsmath}
\usepackage{array}
\usepackage{slashed}
\usepackage{feynmf}
\usepackage{subcaption}
\usepackage{hyperref}
\usepackage[normalem]{ulem}

\def\Lda{\Lambda}

\newcommand{\beq}{\begin{equation}}
\newcommand{\eeq}{\end{equation}}
\newcommand{\bea}{\begin{eqnarray}}
\newcommand{\eea}{\end{eqnarray}}
\newcommand{\beas}{\begin{eqnarray*}}
\newcommand{\eeas}{\end{eqnarray*}}
\newcommand{\bcr}{\begin{center}}
\newcommand{\ecr}{\end{center}}
\def\Re{{\cal R \mskip-4mu \lower.1ex \hbox{\it e}\,}}
\def\Im{{\cal I \mskip-5mu \lower.1ex \hbox{\it m}\,}}

\def\etal{{\it et al.}}

\def\tev{\,{\ifmmode\mathrm {TeV}\else TeV\fi}}
\def\gev{\,{\ifmmode\mathrm {GeV}\else GeV\fi}}
\def\mev{\,{\ifmmode\mathrm {MeV}\else MeV\fi}}
\def\to{\rightarrow}

\def\issue(#1,#2,#3){#1 (#3) #2} 
\def\APP(#1,#2,#3){Acta Phys.\ Polon.\ \issue(#1,#2,#3)}
\def\ARNPS(#1,#2,#3){Ann.\ Rev.\ Nucl.\ Part.\ Sci.\ \issue(#1,#2,#3)}
\def\CPC(#1,#2,#3){comp.\ Phys.\ comm.\ \issue(#1,#2,#3)}
\def\CIP(#1,#2,#3){comput.\ Phys.\ \issue(#1,#2,#3)}
\def\EPJC(#1,#2,#3){Eur.\ Phys.\ J.\ C\ \issue(#1,#2,#3)}
\def\EPJD(#1,#2,#3){Eur.\ Phys.\ J. Direct\ C\ \issue(#1,#2,#3)}
\def\IEEETNS(#1,#2,#3){IEEE Trans.\ Nucl.\ Sci.\ \issue(#1,#2,#3)}
\def\IJMP(#1,#2,#3){Int.\ J.\ Mod.\ Phys. \issue(#1,#2,#3)}
\def\JHEP(#1,#2,#3){J.\ High Energy Physics \issue(#1,#2,#3)}
\def\JPG(#1,#2,#3){J.\ Phys.\ G \issue(#1,#2,#3)}
\def\MPL(#1,#2,#3){Mod.\ Phys.\ Lett.\ \issue(#1,#2,#3)}
\def\NP(#1,#2,#3){Nucl.\ Phys.\ \issue(#1,#2,#3)}
\def\NIM(#1,#2,#3){Nucl.\ Instrum.\ Meth.\ \issue(#1,#2,#3)}
\def\PL(#1,#2,#3){Phys.\ Lett.\ \issue(#1,#2,#3)}
\def\PRD(#1,#2,#3){Phys.\ Rev.\ D \issue(#1,#2,#3)}
\def\PRL(#1,#2,#3){Phys.\ Rev.\ Lett.\ \issue(#1,#2,#3)}
\def\PTP(#1,#2,#3){Progs.\ Theo.\ Phys. \ \issue(#1,#2,#3)}
\def\RMP(#1,#2,#3){Rev.\ Mod.\ Phys.\ \issue(#1,#2,#3)}
\def\SJNP(#1,#2,#3){Sov.\ J. Nucl.\ Phys.\ \issue(#1,#2,#3)}

\title{\LARGE Model-independent Astrophysical Constraints on Leptophilic Dark Matter in the Framework of Tsallis Statistics}

\author[a]{Atanu Guha,}
\author[b]{P. S. Bhupal Dev,}
\author[a,c]{and Prasanta Kumar Das}

\affiliation[a]{Department of Physics, Birla Institute of Technology and Science Pilani, K K Birla Goa Campus, NH-17B, Zuarinagar, Goa 403726, India}

\affiliation[b]{Department of Physics and McDonnell Center for the Space Sciences, Washington University,
St. Louis, MO 63130, USA}

\affiliation[c]{Department of Physics and Astronomy, University of Kansas, Lawrence, KS 66045, USA}

\emailAdd{p20140401@goa.bits-pilani.ac.in}
\emailAdd{bdev@wustl.edu}
\emailAdd{pdas@goa.bits-pilani.ac.in}

\abstract{We derive model-independent astrophysical constraints on leptophilic dark matter (DM), considering its thermal production in a supernova core and taking into account core temperature fluctuations within the framework of $q$-deformed Tsallis statistics. In an effective field theory approach, where the DM  fermions interact with the Standard Model via dimension-six operators of either scalar-pseudoscalar, vector-axial vector, or tensor-axial tensor type, we obtain bounds on the effective cut-off scale $\Lambda$ from supernova cooling and free-streaming of DM from supernova core, and from thermal relic density considerations, depending on the DM mass and the $q$-deformation parameter.  Using  
Raffelt's criterion on the energy loss rate from SN1987A, we obtain a lower bound on $\Lda \gtrsim 3$ (12) TeV corresponding to $q = 1.0~(1.1)$ and an average supernova core temperature of $T_{\rm SN}=30$ MeV.  From the optical depth criterion on 
the free-streaming of DM fermions from the outer 10\% of the SN1987A core, the cooling bound is restricted to $\Lda \gtrsim 1$ TeV. Both cooling and free-streaming bounds are insensitive to the DM mass $m_\chi$ for $m_\chi\lesssim T_{\rm SN}$, whereas for $m_\chi\gg T_{\rm SN}$,  the bounds weaken significantly due to the Boltzmann-suppression of the DM number density. We also calculate the thermal relic density of the DM particles in this setup and find that it imposes an upper bound on $\Lambda^4/m_\chi^2$, which together with the cooling/free-streaming bound significantly constrains light leptophilic DM.  
 }

\begin{document} 
\maketitle
\flushbottom

\section{Introduction} \label{sec:intro}
The existence of a non-relativistic, non-baryonic Dark Matter (DM) component contributing 
to about a quarter of the energy budget of the Universe has been well established by now 
from various astrophysical and cosmological observations~\cite{Bertone:2004pz, Aghanim:2018eyx}. 
However, the nature and properties of DM remain one of the greatest puzzles of modern 
particle physics, despite decades of experimental efforts~\cite{Feng:2010gw}. It is widely 
believed that, in addition to their gravitational interaction, the DM particles should have
other effective couplings to the Standard Model (SM) particles at some level in 
perturbation theory, as e.g., suggested by the Weakly Interacting Massive Particle (WIMP) 
paradigm for thermal DM~\cite{Kolb} and enforced in many beyond SM scenarios~\cite{Arcadi:2017kky}.
However, as highly sensitive searches in both direct and indirect detection experiments, 
as well as collider searches for canonical WIMP DM in the GeV--TeV range have not found 
any signal yet~\cite{PDG}, there is a growing interest in recent years in broadening the 
DM mass range, as well as search strategies~\cite{Battaglieri:2017aum}.  In particular, 
low-mass DM is a very compelling possibility, which might be difficult to find in 
conventional direct detection experiments with nuclear recoil, because the energy 
transferred by the DM particle to the target depends on the reduced mass of the system, 
and the recoil energy for a light DM-nucleon scattering could easily be below the current 
detection threshold of a few keV. On the other hand, if a light DM in the MeV--GeV range 
couples to electrons, the DM scattering with electrons can cause single-electron ionization 
signals, which are detectable with current technology~\cite{Essig:2011nj}. This possibility
has been explored in dedicated experiments, such as XENON10~\cite{Angle:2011th, Essig:2012yx}, 
XENON100~\cite{Aprile:2015ade, Aprile:2016wwo}, DarkSide-50~\cite{Agnes:2018oej},  SENSEI~\cite{Crisler:2018gci} and SuperCDMS~\cite{Agnese:2018col}. Significant improvements in direct detection sensitivity to light DM are expected in the foreseeable future~\cite{Battaglieri:2017aum}. 

There exist complementary, astrophysical bounds on light DM in the MeV--GeV range, as they could be thermally produced in the core of collapsing stars and could subsequently escape the star, causing excessive cooling~\cite{Raffelt:1996wa}. This is particularly relevant for core-collapse type-II supernova explosion~\cite{Mirizzi:2015eza, Janka:2016fox}, such as SN1987A~\cite{Arnett:1990au}, for which 
the total neutrino energy has been estimated, based on the modest quantity of electron anti-neutrinos 
detected~\cite{Hirata:1987hu, Bionta:1987qt, Alekseev:1987ej}. The energy carried away by exotic particles from the supernova core cannot exceed a sizable fraction of the total neutrino energy. Therefore, the SN1987A data can be used to put stringent constraints on light DM couplings with the SM~\cite{Fayet:2006sa, Dreiner:2013mua, Zhang:2014wra, Guha:2015kka, Heurtier:2016otg, Tu:2017dhl, Mahoney:2017jqk, Chang:2018rso, Knapen:2017xzo, Kazanas:2014mca, Kolb:1996pa, Dent:2012mx, Warren:2016slz, Hidaka:2006sg}. However, these constraints were derived assuming the supernova core to be in thermodynamic equilibrium with a fixed core temperature, and therefore, applying either Fermi-Dirac or Bose-Einstein distribution for the particle species. On the other hand, due to ergodicity breaking, the supernova core may remain indefinitely trapped into an out-of-equilibrium quasi-stationary state with a fluctuating core temperature, and therefore, may not follow the Boltzmann-Gibbs equilibrium statistical mechanics with characteristic exponential and Gaussian distributions. Such non-equilibrium behavior is better described by {\it Tsallis statistics}~\cite{Tsallis:1987eu}, obtained by maximizing distributions for the generalized Tsallis entropy $S_q$, which reduces to the Boltzmann-Gibbs entropy as the parameter $q\to 1$. These $q$-Gaussian features have been observed in nature in several astrophysical, geophysical, experimental and model systems~\cite{Tsallis:2009zex}. The Tsallis statistics has also found enormous applications in a variety of other disciplines, including chemistry, biology, geology, mathematics, informatics, economics, engineering, linguistics and others~\cite{Tsallis:2009zex, Sumiyoshi}. In this paper, we revisit the supernova constraints on light DM in the framework of Tsallis statistics. 

 For concreteness, we assume that the DM particles are fermionic and couple exclusively to leptons. Such {\it leptophilic} DM could naturally arise at a more fundamental level in many beyond SM scenarios~\cite{Bernabei:2007gr, Chen:2008dh, Cirelli:2008pk,  Fox:2008kb, Cohen:2009fz, Chun:2009zx, Haba:2010ag, Das:2013jca, Agrawal:2014ufa, Bell:2014tta, Boucenna:2015tra, Duan:2017pkq, Chen:2018vkr, Kopp:2009et, Dev:2013hka, Sui:2018bbh, Dev:2016qbd, Sui:2017qra, Chao:2017emq, Han:2017ars, Lu:2016ups, Madge:2018gfl}, most of which were invoked to address certain experimental anomalies, such as the muon anomalous magnetic moment~\cite{Bennett:2006fi}, DAMA/LIBRA annual modulation~\cite{Bernabei:2013xsa, Bernabei:2018yyw}, anomalous positron excess in ATIC~\cite{Chang:2008aa}, PAMELA~\cite{Adriani:2008zr, Adriani:2013uda}, AMS-02~\cite{Accardo:2014lma, Aguilar:2014mma}, Fermi-LAT~\cite{Abdollahi:2017nat},  DAMPE~\cite{Ambrosi:2017wek} and CALET~\cite{Adriani:2018ktz}, the gamma-ray excess at the galactic center~\cite{TheFermi-LAT:2015kwa}, and the IceCube ultra-high energy neutrino excess~\cite{Aartsen:2017mau, icecube2}. However, we follow a model-independent, effective field theory (EFT) approach~\cite{Kopp:2009et, Chang:2014tea}, assuming that the scale of new physics mediating the DM-SM interactions is heavier than the electroweak scale and that it respects the SM gauge symmetry. Thus, the only relevant degrees of freedom in our analysis are the SM particles, the DM and an effective cut-off scale $\Lambda$ which determines the strength of the four-Fermi operators involving the DM and SM leptons. This EFT approach enables us to eschew the specific details of the underlying new physics models, and to derive model-independent constraints on leptophilic DM from supernova cooling and free-streaming criteria, as well as relic density considerations.

The outline of the paper is  as follows: In section~\ref{sec:tsallis}, we briefly review the framework of 
Tsallis statistics. In Section~\ref{sec:astro}, we discuss the astrophysical bounds on light DM from 
supernova cooling (Section~\ref{sec:raffelt}), free-streaming (Section~\ref{sec:optical}) and relic density 
(Section~\ref{sec:relic}) considerations. In Section~\ref{sec:eft}, we describe our EFT approach and 
derive the pertinent analytic expressions for the energy loss rate, optical depth and relic density 
calculations. In Section~\ref{sec:numerical}, we present our numerical results in terms of constraints on
the EFT cut-off scale $\Lambda$ as a function of the DM mass $m_\chi$ for both undeformed and $q$-deformed 
scenarios in the framework of Tsallis statistics. Our conclusions are given in Section~\ref{sec:conc}. 
In Appendix~\ref{app:q}, we derive an upper bound on $q$ from supernova simulation fits to the neutrino 
spectrum. In Appendix~\ref{app:eff}, we discuss the $q$ and $T$ dependence of the effective degrees of freedom relevant for the relic density constraints. In Appendix~\ref{app:cross}, we give the analytic expressions for the DM production, scattering and annihilation cross sections in our EFT framework. 


\section{Review of Tsallis Statistics} \label{sec:tsallis}
In equilibrium statistical mechanics, the distribution functions are obtained by maximizing the Boltzmann-Gibbs entropy 
\begin{align}
S_{\rm BG} \ = \ k\sum_{i=1}^W p_i \ln \frac{1}{p_i} \, ,
\label{eq:BG}
\end{align}
with the normalization condition $\sum_{i=1}^W p_i = 1$. Here $p_i$ is the probability for the system to be in the $i$-th microstate, $W$ is the total number of microstates, and $k$ is the Boltzmann constant.\footnote{If every microstate has the same probability $p_i=1/W$, Eq.~\eqref{eq:BG} reduces to the famous Boltzmann entropy formula $S=k\ln W$, engraved on his tomb in Vienna.} For a system with equilibrium temperature $T$, 
\begin{align}
p_i = \frac{e^{-\beta \epsilon_i}}{\sum\limits_{i=1}^W e^{-\beta \epsilon_i}} 
\label{eq:pi}
\end{align}
(with $\beta=1/kT$) corresponds to the canonical ensemble probability to observe a microstate of energy $\epsilon_i$. However, if the temperature is fluctuating around an average value and the system is in an out-of-equilibrium quasi-stationary state, the Boltzmann-Gibbs entropy loses its extensive property. It can be generalized to a non-extensive $q$-deformed entropy~\cite{Tsallis:1987eu} 
\begin{align}
S_q \ = \ k\sum_{i=1}^W p_i \ln_q \frac{1}{p_i} \ \equiv \ \frac{k\left(1-\sum\limits_{i=1}^W p_i^q\right)}{q-1} \, ,
\end{align}
such that in the limit $q\to 1$, it coincides with the Boltzmann-Gibbs entropy, i.e. $S_1=S_{\rm BG}$. This feature allows to describe these special non-equilibrium states with the same formal framework of the equilibrium statistical mechanics, known as the Tsallis statistics.

Extremizing $S_q$ subject to constraints yields a generalized canonical 
ensemble where the probability to observe a microstate of energy $\epsilon_i$ is given by 
\bea
p_i \ = \ e_q^{-\beta \epsilon_i} \ \equiv \ \frac{1}{\left[ 1 + (q - 1)\beta \epsilon_i\right]^{\frac{1}{q - 1}}} \, ,
\label{eq:piq}
\eea
which is a generalization of Eq.~\eqref{eq:pi}. The $q$-deformed probability distribution function may be driven  by the non-equilibrium situation with local temperature fluctuation~\cite{Beck_cohen} over a region of space, which can be accounted for by defining a $\chi^2$ distribution of the form~\cite{Kaniadakis}     
\bea
f(\beta) \ = \ \frac{1}{\Gamma\left(\frac{n}{2}\right)} \left(\frac{n}{2 \beta_0} \right)^{n/2} \beta^{\frac{n}{2} - 1} ~\exp\left(- \frac{n \beta}{2 \beta_0}\right) \, ,
\label{fbeta}
\eea 
where $n$ is the degree of the distribution, i.e. the number of independent  Gaussian random variables $X_i,~i = 1,....,n$ 
and $\beta = \sum_{i=1}^{n} X_i^2$ is the fluctuating inverse temperature, with the average value 
\begin{align}
\beta_0 \ \equiv \ \langle \beta \rangle \ = \ n\langle X_i^2\rangle \ = \ \int\limits _0^\infty d\beta \:  \beta f(\beta) \, .
\end{align} 
Taking into account the local temperature fluctuation, integrating over all $\beta$, we find
the $q$-generalized Maxwell-Boltzmann distribution
\bea
{\mathcal{P}}(E) \ \equiv \ \frac{1}{Z} B(E) \ = \ \frac{1}{Z}\int_{0}^{\infty} d\beta \: e^{-\beta E} f(\beta)
\ = \ \frac{1}{\left[1 + (q-1)bE\right]^{\frac{1}{q-1}}} \ \equiv \ e_q^{- \beta_0 E} \, ,
\eea
where  $q = 1 + \frac{2}{n + 6}$ and $b = \frac{\beta_0}{4 - 3 q}$ and the normalization constant $Z = \int_{0}^{\infty} B(E) dE$.

The generalization of the Maxwell-Boltzmann distribution to Fermi-Dirac and Bose-Einstein distributions is worked out in Ref.~\cite{Beck:2009uy}. The average occupation number of any particle within  this $q$-deformed statistics formalism is given by 
\bea
f(\beta,E) \ = \ \frac{1}{\left[1 + (q-1) bE \right]^{\frac{1}{q-1}} \pm 1} \ \equiv \ \frac{1}{e_q^{\beta_0 E}\pm 1} \, ,
\label{eq:fq}
\eea
where the $+$ ($-$) sign is for fermions (bosons) and $e_q^{\beta_0 E}$ is the effective Boltzmann factor (with $E=\sqrt{\bm{p}^2+m^2}$ being the relativistic energy and $\bm{p}$ being the 3-momentum). Note that in the $q \to 1$ limit, $e_q^{\beta_0 E}$ reduces to the usual Boltzmann factor $e^{-bE}=e^{-\beta_0E}$ (see Appendix A of Ref.~\cite{Guha:2015kka}). For a particle with non-zero chemical potential $\mu$, we can just replace the energy $E$ in Eq.~\eqref{eq:fq} by $\bar{E}\equiv E-\mu$. 

\section{Astrophysical Bounds on Dark Matter }\label{sec:astro}
The idea of putting an astrophysical bound on new particles (e.g.~DM) is simple. If they are 
found to be light, they may be produced copiously inside the astrophysical object and can escape the object, taking away part of its energy and causing excessive cooling. This may contradict the standard theoretical model of cooling of the astrophysical object (e.g. supernova) and its experimental observation. SN1987A provides one of the most powerful natural laboratories for this purpose due to its high density, high temperature and proximity to Earth. Below we briefly discuss the SN1987A cooling and its energy loss rate criterion (see Section~\ref{sec:raffelt}). We also discuss the optical depth criterion for free-streaming of DM particles from SN1987A (see Section~\ref{sec:optical}). Besides these, we also outline the computation of the relic density constraint of the DM in this scenario (see Section~\ref{sec:relic}).

\subsection{Supernova Cooling and Raffelt's Criterion} \label{sec:raffelt}

The supernova SN1987A~\cite{Arnett:1990au}, the most evident example of a core-collapse type II supernova 
explosion known till date, released an enormous amount of energy
which equals to the gravitational binding energy $E_g$, given by
\beq
E_g \ = \ \frac{3 G_N M_{\rm PNS}^2}{5 R_{\rm PNS}} \ \sim \ 3 \times 10^{53} \;{\rm erg} \, ,
\eeq
where $G_N$ is Newton's gravitational 
constant, $M_{\rm PNS}= 1.5 M_\odot$ ($M_\odot$ being the solar mass) is the mass and $R_{\rm PNS} = 10~{\rm km}$ is the radius of the proto-neutron star (PNS). According to our current understanding, neutrinos, produced in supernova explosion, carry away $99\%$ of the released energy, while the remaining $1\%$ contributes to the kinetic energy of the explosion. For the earth-based detectors, the primary astrophysical interest was to detect this neutrino burst. The SN1987A neutrino flux was detected by three experiments, Kamiokande II~\cite{Hirata:1987hu}, IMB~\cite{Bionta:1987qt} and Baksan~\cite{Alekseev:1987ej}, using their earth-based detectors, which detected 12, 8 and 5 antineutrinos, respectively. The data obtained by them suggests that,  in less than 13 seconds, about $10^{53}~\rm{erg}$ energy was released in 
the SN1987A explosion. The observed neutrino luminosity in each detector is $L_\nu \sim 2 \times 10^{53}\;{\rm erg ~s^{-1}}$ (including $3$ generations of neutrinos and anti-neutrinos 
i.e. $\nu_e,~\nu_\mu,~\nu_\tau$ and ${\bar{\nu}_e},~{\bar{\nu}_\mu},~{\bar{\nu}_\tau}$). 
So the observed luminosity per neutrino is 
$\tilde{L}_\nu = \frac{L_\nu}{6} \sim 3 \times 10^{52}~{\rm erg~ s^{-1}}$ and 
the average energy 
loss rate per unit mass is 
$\dot{\varepsilon}\simeq \tilde{L}_\nu/M_{\rm PNS} \simeq 10^{19}~{\rm erg ~g^{-1}s^{-1} }$, which is the energy carried away by each of the above 6 (anti)neutrino species~\cite{Janka:2006fh}. 

 Now besides neutrinos, light particles like axion, Kaluza-Klein graviton, neutralino or DM, if produced inside the SN1987A core, can take away part of the energy released in SN1987A explosion. According to Raffelt's  criterion~\cite{Raffelt:1996wa}, the energy loss rate for these new channels should be less than the above-mentioned average energy loss rate, i.e. 
\beq
\dot{\varepsilon}_{\rm new} \ \le \ 10^{19}~\rm{erg~gm^{-1}~s^{-1}} \, .
\label{eq:upper}
\eeq
This is because any new channel with an emissivity greater than this value 
will take away excessive energy and thus invalidate the observational data. Now in a realistic scenario,
since the core temperature of the supernova is fluctuating,  we will work within the formalism of Tsallis 
statistics~\cite{Tsallis:2009zex} and compute the emissivity as a function of the $q$-parameter, as well 
as of the DM mass and effective cut-off scale (see Section~\ref{sec:eft}). 
Note that the $q$-deformed distribution function \eqref{eq:fq} leads to a modified neutrino spectrum from 
the supernova core, which is however consistent with the current state-of-the-art supernova simulation 
fits~\cite{Tamborra:2012ac, Nikrant:2017nya} as long as $q\leq 1.27$ (see Appendix~\ref{app:q} for details).
Also note that the upper bound \eqref{eq:upper} on $\dot{\varepsilon}_{\text {new}}$ is a data-driven entity and it still valid within the Tsallis statistics framework. For our leptophilic DM scenario, the dominant production channel of DM in the supernova core is the electron-positron annihilation: $e^+e^-\to \chi \bar{\chi}$, where $\chi$ stands for the DM. 

\subsection{Free-streaming of Dark Matter from Supernova Core} \label{sec:optical}
The constraint derived on the EFT cut-off scale $\Lambda$ from the Raffelt's energy loss criterion makes sense if the DM particles produced inside the supernova core always free-stream out carrying away the energy. The free-streaming/trapping length of the DM particles can be estimated in terms of   its mean-free path $\lambda_{\chi}$, which can be evaluated as 
\bea
\lambda_{\chi} \ = \ \frac{1}{n_\psi \cdot \sigma_{\psi \chi \rightarrow \psi \chi}} \, ,
\protect\label{mean_free_path}
\eea
where $\psi$ corresponds to electron ($e$) or nucleon ($N$), $n_\psi$ (with $\psi = e, N$) corresponds to the number density of the target electrons or nucleons in the supernova core 
and $ \sigma_{\psi \chi \rightarrow \psi \chi}$ is the cross-section for the scattering of the 
DM fermion on the target electron or nucleon. In the case of supernova cooling due to light DM, nucleon-DM scattering will be negligible for free-streaming due to the nucleon mass. Moreover, we are considering a leptophilic DM, so any possible DM-nucleon coupling can only arise at the loop level~\cite{Kopp:2009et}. Therefore, we will only focus on the electron-DM scattering: $e\chi\to e\chi$. To obtain constraints on DM from free-streaming, 
we use the optical depth criterion~\cite{Shapiro:1983, Raffelt:2001kv, Janka:2017vlw, Burrows:1986me}
\footnote{If the opacity of particles is dominated by true absorption processes,
the emergent particles  originate near and above the layer at which 
optical depth $\tau\approx 2/3$~\cite{Shapiro:1983}.}  
 \bea \label{Eqn:free-streaming}
 \int_{r_0}^{R_c}\frac{dr}{\lambda_{\chi}} \ \leq \ \frac{2}{3} \, ,
 \eea
 to find out whether the DM produced at a depth $r_0$ free-streams out of the supernova core of radius $R_c$. 
Note that most of the DM production from electron-positron annihilation occurs in the outermost $10\%$ of the supernova core~\cite{Dreiner:2003wh, Dreiner:2013mua}. This is because the outer region has the highest temperature and the lowest electron degeneracy. Hence, we set $r_0 = 0.9~R_c$ in Eq.~\eqref{Eqn:free-streaming} to derive the free-streaming bound on the effective cut-off scale $\Lambda$ as a function of the DM mass (see Section~\ref{sec:numerical}).  
\subsection{Relic Density} \label{sec:relic}
Any stable species should contribute to the overall energy density of the universe. In the WIMP DM scenario, the DM particles are in thermal (and chemical) equilibrium with the cosmic plasma at high temperatures and get decoupled (freeze-out) as the plasma temperature drops below the DM mass due to Hubble expansion of the universe. In our leptophilic DM setup, the same interactions that produce the DM particles in the supernova core through electron-positron annihilation would also help the DM particles annihilate back into electron-positron pairs: $\chi\bar{\chi}\to e^+e^-$, the rate of which then sets their present-day thermal relic abundance, given by the standard expression~\cite{Gondolo:1990dk}  
\bea 
\Omega_{\chi} h^2 \ = \ 2.755 \times 10^{8}\: Y_0 \left(\frac{m_{\chi}}{1~\rm GeV}\right) 
\label{Eqn:relic-density}
\eea
where $\Omega_\chi\equiv \rho_\chi/\rho_{\rm crit}$ is the ratio of the DM density and critical density of the universe, $h=0.678 \pm 0.009$ is the scale factor for Hubble expansion rate~\cite{PDG}, $m_{\chi}$ is the DM mass, and $Y_0$ is the present-day value of the yield $Y=n_\chi/s$ ($n_\chi$ being the DM number density and $s$ the total entropy density) at present temperature of the universe $T_0=2.726~{\rm K}$, which is obtained by solving the Boltzmann equation~\cite{Gondolo:1990dk}:   
\bea
\frac{1}{Y_0} \ = \ \frac{1}{Y_f} + \left(\frac{45G_N}{\pi} \right)^{-\frac{1}{2}} 
\int\limits_{T_0}^{T_f} g_{*}^{\frac{1}{2}}(T) \langle \sigma v_{\rm rel}\rangle dT \, .
\label{eqn:Y0}
\eea
Here $Y_f$ is the value at the freeze-out temperature $T=T_f$, $g_* (T)$ is the total effective relativistic degrees of freedom at temperature $T$ and $\langle \sigma v_{\rm rel}\rangle$ is the thermal-averaged annihilation cross section times the relative velocity between the two annihilating DM particles: 
\bea
\langle \sigma v_{\rm rel}\rangle \ = \ \frac{\int \sigma v_{\rm rel} \: e^{-E_1/T} e^{-E_2/T} d^3 \bm{p}_1 d^3 \bm{p}_2}{\int e^{-E_1/T} e^{-E_2/T} d^3 \bm{p}_1 d^3 \bm{p}_2} \, .
\label{eq:sigmav}
\eea
Matching Eq.~\eqref{Eqn:relic-density} to the observed value of non-baryonic, cold DM density 
$\Omega^{\rm obs}_{\rm{CDM}} h^2=0.1186\pm 0.0020$~\cite{PDG}, we derive the relic density constraint on the effective cut-off scale $\Lambda$ as a function of the DM mass $m_\chi$ in Section~\ref{sec:numerical}. The $q$-dependence of the relic density constraint~\cite{Guha:2018yzz} comes from the thermal averaging, as well as due to the total effective degrees of freedom
$g_{*}$ which depends on the temperature $T$ and the deformation parameter $q$. See Appendix~\ref{app:eff} for a detailed discussion on $g_{*}(T)$ in the $q$-deformed scenario.  

\section{Effective Operator Approach} \label{sec:eft}
\begin{figure}[t!]
  \centering
 \includegraphics[width=0.35\textwidth]{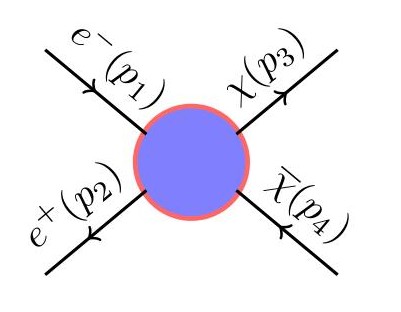}
   \caption{Feynman diagram for the effective four-Fermi interactions induced by the effective Lagrangian \eqref{eq:eft}. Depending on the arrow of time, this represents $ e^-e^+ \to \chi \bar{\chi}$ (relevant for supernova cooling), $ e^\pm \chi \to e^\pm\chi$ (free-streaming) or $\chi\bar{\chi}\to e^-e^+$ (relic density). Here $p_i$'s (with $i=1,\cdots, 4$) denote the 4-momenta of the incoming and outgoing particles.}
\label{feynman}
\end{figure}

We perform a model-independent analysis for the astrophysical constraints working under the assumption that the DM 
fermion $\chi$ is leptophilic, i.e. couples directly only to the SM leptons. The specific processes of our interest are (i) $e^- e^+ \to \chi \bar{\chi}$ (for SN1987A cooling), (ii) $e^\pm \chi \to e^\pm \chi$ (for free-streaming), and (iii) $\chi \bar{\chi} \to e^- e^+$ (for relic density), which are all related by crossing symmetry, as shown in Fig.~\ref{feynman}. 
The most general effective Lagrangian leading to these processes is given by the following dimension-six four-Fermi operators~\cite{Kopp:2009et}:
\bea 
\mathcal{L}_{\rm eff} \ = \ \frac{1}{\Lambda^2}\sum_j  \left(\bar{\chi} \Gamma^j_{\chi} \chi \right) \left(\bar{e} \Gamma^j_{e} e \right) \, ,
\label{eq:eft}
\eea
where $\Lambda$ is the cut-off scale for the EFT description and the index $j$ corresponds to different Lorentz structures,  such as scalar (S),  pseudo-scalar (P), vector (V), axial-vector (A), tensor (T) and axial-tensor (AT) currents. We classify them as follows: 
\begin{align}
&\text{S-P type}: & &
\Gamma_{\chi} \ = \ c^{\chi}_{S}+i c^{\chi}_{P} \gamma_5 \, ,  & & \Gamma_{e} = c^{e}_{S}+i c^{e}_{P} \gamma_5 \, , \nonumber \\
&\text{V-A type}: & &
\Gamma_{\chi}^{\mu} = \left( c^{\chi}_{V}+ c^{\chi}_{A} \gamma_5 \right) \gamma^{\mu} \, ,  & &
\Gamma_{e\mu} = \left( c^{e}_{V}+ c^{e}_{A} \gamma_5 \right) \gamma_{\mu} \, , \nonumber \\
&\text{T-AT type}: & &
\Gamma_{\chi}^{\mu \nu} = \left( c^{\chi}_{T}+i c^{\chi}_{AT} \gamma_5 \right) \sigma^{\mu \nu} \, , & & \Gamma_{e \mu \nu} = \sigma_{\mu \nu} \, , 
\label{eq:op}
\end{align}
where $\sigma^{\mu \nu}=\frac{i}{2}[\gamma^\mu,\gamma^\nu] $ is the spin tensor\footnote{We do not write an AT part for $\Gamma_{e\mu\nu}$, because the AT$\otimes$AT coupling is equivalent to T$\otimes$T, and similarly, T$\otimes$AT is equivalent to AT$\otimes$T due to the identity $\sigma^{\mu\nu}\gamma_5=\frac{i}{2}\epsilon^{\mu\nu\alpha\beta}\sigma_{\alpha\beta}$.} and $c_j^{\chi,e}$ are dimensionless, real couplings. For simplicity, we use a common cut-off scale in Eq.~\eqref{eq:eft} for all these operators. 
Since we are making a model-independent analysis, we do not discuss any specific realization of the above set of effective operators. 
We investigate SN1987A cooling, DM free-streaming  and relic density constraints in light of each of the operator types listed above, taken one at a time, and obtain the corresponding bounds on the cut-off scale $\Lambda$ as a function of the DM mass. Within the Tsallis statistics framework, we will consider the general case with $q\neq 1$, as well as the Boltzmann-Gibbs limit of $q=1$.   

\subsection{SN1987A Cooling} 
Electrons are abundant in the supernova core. At high temperatures, positrons are also present, since after $\sim 10-20$ ms of the burst, their production is no longer inhibited~\cite{Janka:2006fh, Bethe:1990mw}. These electron-positron pairs can interact via the leptophilic effective 
operators \eqref{eq:eft} to pair-produce light DM particles: $ e^-(p_1)e^+(p_2) \stackrel{}{\longrightarrow} \chi(p_3)\bar{\chi}(p_4) $, as shown in Fig.~\ref{feynman}.  

Once produced, the DM particles can take away a fraction of the energy released in the supernova explosion and contribute to the energy loss rate~\cite{Raffelt:1996wa}. Within the framework of Tsallis statistics, the energy loss rate is given by~\cite{Guha:2015kka}
\bea \label{eqn:energy-loss-rate}
\dot{\varepsilon}_{e^- e^+ \rightarrow \chi \bar{\chi}}(q) 
& \ = \ & \frac{1}{\rho_{\rm SN}} \langle n_{e^-} n_{e^+} \sigma_{e^- e^+ \rightarrow \chi \bar{\chi}}~ v_{\rm rel} E_{\rm cm}\rangle \nonumber \\
 &\ = \ & \frac{1}{2\pi^4\rho_{\rm SN}}\int\limits_{m_\chi}^{\infty} \int\limits_{m_\chi}^{\infty} d E_1 d E_2  \: E_1 E_2 (E_1 + E_2)^3~\sigma_{e^- e^+ \rightarrow \chi \bar{\chi}} \: f_1f_2
\eea 
where $ E_{\rm cm}=E_1+E_2\equiv \sqrt s$ is the center-of-mass energy ($E_{1,2}$ being the electron and positron energies respectively, which are taken equal here), $ v_{\rm rel}=s/4E_1 E_2$ is the relative velocity between the colliding electron and positron, and $\rho_{\rm SN}$ is the supernova matter density, $f_{1}, f_2$  are the distribution functions [cf.~Eq.~\eqref{eq:fq}]  and $n_{e^-}= 2\int \frac{d^3\bm{p}_1}{(2\pi)^3}f_1$ and $n_{e^+}= 2\int \frac{d^3\bm{p}_2}{(2\pi)^3}f_2$ are the number densities for electron and positron, respectively. The analytic expressions for the cross sections $\sigma_{e^- e^+ \rightarrow \chi \bar{\chi}}$ for different four-Fermi operators listed in Eqs.~\eqref{eq:op} are given in Appendix~\ref{app:cross1}. We will present our numerical results for the simplified case when all the effective coupling parameters in Eq.~\eqref{eq:eft} are assumed to be the same and normalized to unity, i.e.
$c_S^e=c_P^e=c_S^{\chi}=c_P^{\chi}=1$ (S-P type), $c_V^e=c_A^e=c_V^{\chi}=c_A^{\chi}=1$ (V-A type)
and $c_T^{\chi}=c_{AT}^{\chi}=1$ (T-AT type).\footnote{This assumption is not valid for a Majorana DM, for which the vectorial and tensorial interactions are forbidden, i.e. $c_V^\chi=c_T^\chi=c_{AT}^\chi=0$. So the bounds derived in Section~\ref{sec:numerical} will have to be reinterpreted accordingly in that scenario.} Given the general expressions for the cross sections in Appendix~\ref{app:cross}, our results can be easily translated to other coupling choices. In the simplified case, the expressions \eqref{eq:A1}-\eqref{eq:A3} can be reduced to  the following:
\bea 
\sigma^{\rm S-P}_{e^-e^+ \to \chi \bar{\chi}} \ & = & \ \frac{A}{4 \pi  \Lambda ^4 s} \left[ s^2 -2 s (m_\chi^2 + m_e^2) + 4 m_\chi^2 m_e^2\right] \label{eqn:totalcross-section-cooling-scalar}, \\
\sigma^{\rm V-A}_{e^-e^+ \to \chi \bar{\chi}} \ & = & \ \frac{A}{3 \pi  \Lambda ^4 s} \left[ s^2 - s (m_\chi^2 + m_e^2) 
+ 4 m_\chi^2 m_e^2 \right] \, , \label{eqn:totalcross-section-cooling-vector} \\
\sigma^{\rm T-AT}_{e^-e^+ \to \chi \bar{\chi}} \ & = & \ \frac{A}{3 \pi  \Lambda ^4 s} \left[ s^2 + 2 s (m_e^2 + m_\chi^2) + 
4 m_{\chi}^2 m_e^2 \right] \, , \label{eqn:totalcross-section-cooling-tensor}
\eea
where $A = \sqrt{(s-4 m_{\chi}^2)/(s-4 m_{e}^2)}$. 


Introducing the dimensionless variables $x_i = E_i/kT$~(with $i=1,2$), we can write down the energy 
loss rate in Eq.~(\ref{eqn:energy-loss-rate}) as 
\begin{align} 
\dot{\varepsilon}_{e^- e^+ \to \chi \bar{\chi}} (q) \ & = \ \frac{T^{7}}{2\pi^4\rho_{\rm SN}} \int\limits_{\frac{m_\chi}{T}}^{\infty}
\int\limits_{\frac{m_\chi}{T}}^{\infty} dx_1dx_2 \nonumber \\
& \qquad \qquad \times \frac{x_1x_2(x_1+x_2)^3 \: \sigma_{e^- e^+ \rightarrow \chi \bar{\chi}}}
{\left(\left[1 + \frac{bT}{\tau} \left(x_1 - \frac{\mu_{e^-}}{T}\right) \right]^{\tau} + 1\right)\left(\left[1 + \frac{bT}{\tau} \left(x_2 - \frac{\mu_{e^+}}{T}\right)\right]^{\tau} + 1\right)} \, ,
\label{eq:eps1}
\end{align}
where $\tau=1/(q-1)$. 
In the supernova core, the electrons and positrons are assumed to be in chemical equilibrium, so $\mu_{e^+}=-\mu_{e^-}$.  In the $q \to 1$ limit, the $q$-deformed distribution in Eq.~\eqref{eq:eps1} approaches the usual Fermi-Dirac distribution, and therefore, the energy loss rate in the undeformed scenario becomes 
the energy loss rate in the $q=1$ case takes the following form 
\bea \label{eqn:energy-loss-rate-undeformed}
\dot{\varepsilon}_{e^- e^+ \to \chi \bar{\chi}}(q=1)\ = \ \frac{ T^{7}}{2\pi^4\rho_{\rm SN}} 
\int\limits_{\frac{m_\chi}{T}}^{\infty} \int\limits_{\frac{m_\chi}{T}}^{\infty} dx_1dx_2 \frac{x_1x_2(x_1+x_2)^3 \: \sigma_{e^- e^+ \rightarrow \chi \bar{\chi}}}
{\left[\exp {\left(x_1- \frac{\mu_{e^-}}{T} \right)}  + 1\right]\left[\exp {\left(x_2- \frac{\mu_{e^+}}{T} \right)}  + 1\right]} \nonumber \, . \\
\label{eq:eps2}
\eea 
We will use Eqs.~\eqref{eq:eps1} and \eqref{eq:eps2} in our numerical analysis (see Section~\ref{sec:numerical}) to compare the energy loss constraints in the $q$-deformed and undeformed scenarios. 

\subsection{Free-Streaming} 
As discussed in Section~\ref{sec:optical}, the free-streaming of DM from the supernova core depends on the mean free-path $\lambda_\chi$ of the DM fermion. Given the EFT Lagrangian (\ref{eq:eft}), the DM mean free-path is governed by the scattering process $e^\pm \chi \to e^\pm \chi$ and $\lambda_\chi$ is inversely proportional to the cross-section $\sigma_{e \chi \to e \chi}$ [cf.~Eq.~\eqref{mean_free_path}].  For different four-Fermi operators listed in Eqs.~\eqref{eq:op}, the relevant cross sections are given in Appendix~\ref{app:cross2}. Under the simplifying assumption that all the effective coupling parameters in Eq.~\eqref{eq:eft} are the same and equal to unity, we can rewrite Eqs.~\eqref{eq:A4}-\eqref{eq:A6} as   
\bea
 \sigma^{\rm S-P}_{e \chi \to e  \chi } \ & = & \ \frac{1 }{12 \pi \Lambda^4  s^3 } \left[ s^4 - m_{\chi}^2 s^3 - m_{\chi}^6 s + m_{\chi}^8 \right] \, , \label{eq:free1} \\
 \sigma^{\rm V-A}_{e \chi \to e  \chi } \ & = & \ \frac{1 }{\pi \Lambda^4  s } \left[s^2  - 2 m_{\chi}^2 s + m_{\chi}^4 \right] \, , \\
 \sigma^{\rm T-AT}_{e \chi \to e  \chi } \ & = & \ \frac{1 }{3 \pi \Lambda^4  s^3 }
\left[ 7 s^4 - 13 m_{\chi}^2 s^3 + 6 m_{\chi}^4 s^2 - m_{\chi}^6 s 
 +  m_{\chi}^8 \right] \, ,\label{eq:free3}
 \eea
where $s=(E_e+E_\chi)^2$, where $E_e=\sqrt{m_e^2+(kT)^2}$ and $E_\chi=\sqrt{m_\chi^2+(kT)^2}$. 
We will use the optical depth criterion \eqref{Eqn:free-streaming} with the mean free path given by Eq.~\eqref{mean_free_path} and the cross sections given above to derive the free-streaming bounds in Section~\ref{sec:numerical}. Note that the mean free-path as such does not depend on the $q$-parameter in the framework of Tsallis statistics. However, as the DM mass exceeds the average supernova core temperature, we should multiply $\lambda_\chi^{-1}$ with the effective Boltzmann suppression factor $e_q^{-\beta_0E}$ [cf.~Eq.~\eqref{eq:fq}] to take into account the decreasing DM number density in the core, which induces a mild $q$-dependence, as we will see in Section~\ref{sec:numerical}. 
\subsection{Relic Density} 
The thermal relic density of DM [cf.~Eq.~\eqref{Eqn:relic-density}] depends on the thermal average
of the annihilation cross-section $\sigma_{\chi \bar{\chi} \to e^- e^+}$ times the relative
velocity $v_{\rm rel}$ of the two colliding DM fermions produced in early era. For different four-Fermi operators listed in Eqs.~\eqref{eq:op}, the relevant cross sections are given in Appendix~\ref{app:cross3}. Under the simplifying assumption that all the effective coupling parameters in Eq.~\eqref{eq:eft} are the same and equal to unity, the annihilation cross-sections in Eqs.~\eqref{eq:A7}-\eqref{eq:A9} can be rewritten as follows:
\bea \label{eqn:totalcross-section-relic-scalar}
\sigma^{\rm S-P}_{\chi \bar{\chi} \to e^-e^+ } \ & = & \ \frac{1}{4 A \pi  \Lambda ^4 s} \left[ s^2 -2 s (m_\chi^2 + m_e^2) + 4 m_\chi^2 m_e^2\right] \, ,\\
\sigma^{\rm V-A}_{\chi \bar{\chi} \to e^-e^+} \ & = & \ \frac{1}{3 A \pi \Lambda ^4 s} \left[ s^2 - s (m_\chi^2 + m_e^2) 
+ 4 m_\chi^2 m_e^2 \right] \, ,\label{eqn:totalcross-section-relic-vector}\\
\sigma^{\rm T-AT}_{\chi \bar{\chi} \to e^-e^+} \ & = & \ \frac{1}{3 A \pi  \Lambda ^4 s} \left[ s^2 + 2 s (m_e^2 + m_\chi^2) + 
4 m_{\chi}^2 m_e^2 \right] \, ,\label{eqn:totalcross-section-relic-tensor}
\eea
where $A = \sqrt{(s-4 m_{\chi}^2)/(s-4 m_{e}^2)}$ and $s=(E_1+E_2)^2$ (with $E_{1,2}$ being the energies of $\chi$ and $\bar{\chi}$, respectively). 
Also from Eq.~\eqref{eq:sigmav}, the thermal averaged cross-section times velocity for the DM fermion pair-annihilation is given by 
\bea \label{eqn:thermal_averaged_crosssection_times_velocity}
\langle \sigma_{\chi \bar{\chi} \rightarrow e^- e^+} v_{\rm rel}\rangle \ = \  
\frac{\int\limits_{m_\chi}^{\infty} \int\limits_{m_\chi}^{\infty} d E_1d E_2 \: E_1E_2(E_1 + E_2)^2
\sigma_{\chi \bar{\chi} \rightarrow e^- e^+}f_1f_2}{\int\limits_{m_\chi}^{\infty} \int\limits_{m_\chi}^{\infty} d E_1d E_2 \: E_1^2E_2^2f_1f_2} \,
\eea
where $f_{1,2}$ are the distribution functions for $\chi$ and $\bar{\chi}$ as given by Eq.~\eqref{eq:fq}. We will use Eq.~\eqref{eqn:thermal_averaged_crosssection_times_velocity} to calculate the relic density constraints in the following section. 
\section{Numerical Results} \label{sec:numerical}
In this section, we use the Tsallis statistics framework [cf.~Section~\ref{sec:tsallis}] 
and EFT approach [cf.~Section~\ref{sec:eft}] to derive the cooling, free-streaming 
and relic density bounds on the leptophilic DM scenario. For the $q$-deformed 
distribution function given by Eq.~\eqref{eq:fq} and the energy loss rate given 
by Eq.~\eqref{eq:eps2}, we use the average supernova core temperature 
of $T=T_{\rm SN}=30~\rm{MeV}$. Similarly, in Eq.~\eqref{Eqn:free-streaming}, we 
use $R_c=10$ km, in Eq.~\eqref{eq:eps2}, we use the supernova matter density of
$\rho_{\rm SN}=3\times 10^{14}~{\rm gm.cm}^{-3}$ and chemical potential of 
$\mu_{e^-}=-\mu_{e^+}=200~{\rm MeV}$, and in Eq.~\eqref{mean_free_path}, we 
use the electron number density of $n_e=10^{37}~{\rm cm}^{-3}$ ~
\cite{Raffelt:1996wa, Burrows:1988ah}. For the deformation parameter $q$,
we choose a benchmark value of $q=1.1$~\footnote{This is consistent with the upper bound on $q$ from the 
numerical fits to the supernova neutrino spectrum (see Appendix~\ref{app:q}).} and compare our results with the undeformed 
scenario with $q=1$. As for the effective couplings of DM to electrons given by Eqs.~\eqref{eq:op}, we assume a simplified case with only one type of interaction at a time and all corresponding couplings being equal and of order unity. The relevant cross section expressions given in Appendix~\ref{app:cross} greatly simplify in this case, as already discussed in the previous Section. 

\begin{figure}[t!]
\centering
\includegraphics[width=0.49\textwidth]{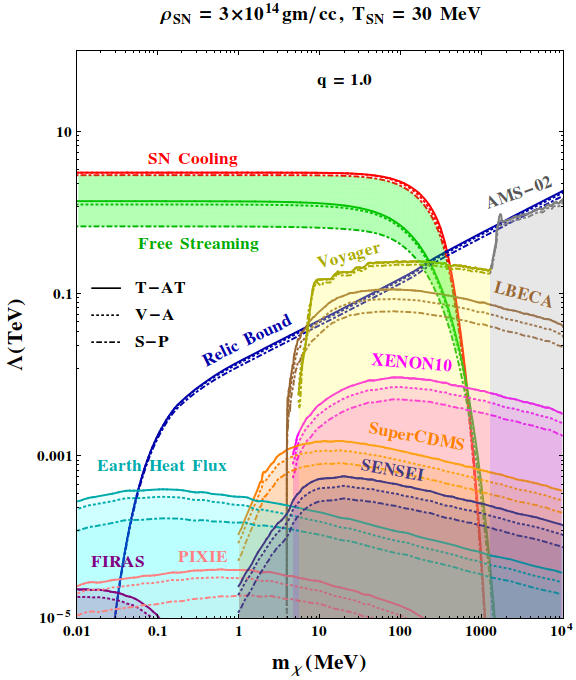}
\includegraphics[width=0.49\textwidth]{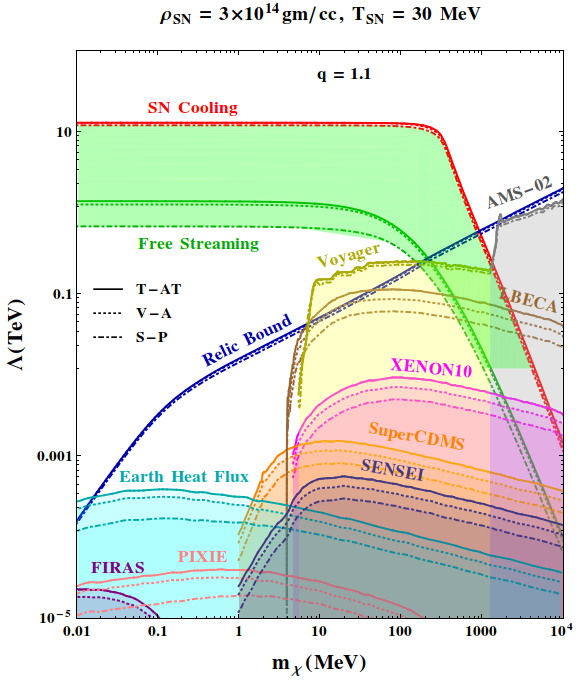}
\caption{The bound on $\Lda$ obtained from SN1987A 
cooling (red) and free-streaming (green) are plotted against the DM mass in 
undeformed ($q = 1.0$, left panel) and deformed ($q=1.1$, right panel) scenarios within the framework of 
Tsallis statistics. The green shaded regions are excluded from SN cooling and free-streaming 
bound. The blue lines indicate the values of $\Lambda$ for a given DM mass to satisfy the relic density 
constraint and the regions above the blue lines are excluded from overclosure constraint. The solid, 
dashed and dot-dashed lines in each case correspond to the T-AT type, V-A type and S-P type interactions 
[cf.~Eqs.~\eqref{eq:op}], respectively. The shaded magenta, pink and violet regions are excluded from 
the direct detection experiments XENON10~\cite{Essig:2012yx}, SuperCDMS~\cite{Agnese:2018col} and 
SENSEI~\cite{Crisler:2018gci} data, respectively. On the other hand, the gray,  yellow shaded regions are 
ruled out by the indirect detection experiments AMS-02 and Voyager1 data~\cite{Boudaud:2016mos}, 
respectively. The cyan shaded region are forbidden due to the constraints from the internal heat flux 
of earth~\cite{Chauhan:2016joa}. The purple shaded regions are excluded by CMB spectral distortion data 
from FIRAS~\cite{Ali-Haimoud:2015pwa}. The brown and pink curves are the projected limits from 
LBECA~\cite{LBECA} and PIXIE~\cite{Ali-Haimoud:2015pwa} respectively. The supernova parameters used in this analysis are:  $T_{\rm SN}=30~{\rm MeV},~\rho_{\rm SN}=3\times 10^{14}~\rm{gm\: cm}^{-3}$ and $n_e=10^{37}~\rm{cm}^{-3}$.}
\label{fig:lambda-mchi1-T30-q1.0}
\end{figure}

We present our results in the plane of DM mass $m_\chi$ and the effective cut-off scale $\Lambda$. 
The DM mass is varied between 10 keV--10 GeV.  The lower value of the DM mass range is chosen to be 
consistent with the generic lower bound of $\sim$ keV on fermion DM, based on DM phase space density 
distribution in dwarf spheroidal galaxies~\cite{Tremaine:1979we, Boyarsky:2008ju}. As for the upper value 
of the mass range chosen here, this will be justified below, where we show that for $m_\chi\gg T_{\rm SN}$,
the supernova constraints become weaker than other existing constraints. Our results are shown 
in Fig.~\ref{fig:lambda-mchi1-T30-q1.0} for both undeformed ($q=1$, left panel) and deformed ($q=1.1$, 
right panel) scenarios. In each case, the solid, dashed and dot-dashed lines correspond to the T-AT type, 
V-A type and S-P type interactions [cf.~Eqs.~\eqref{eq:op}], respectively, taken one at a time and setting 
the other interactions to zero. Although we show the constraints on the EFT scale
$\Lambda$ all the way down to 10 MeV, one should be careful while applying these constraints to heavier DM.
This is because the EFT approach is strictly valid only if the cut-off scale $\Lambda$ is above the 
DM mass scale.

\subsection{Supernova Cooling Bound} 
The supernova cooling bounds (red lines) are obtained by using the Raffelt criterion [cf.~Section~\ref{sec:raffelt}], i.e. by requiring the energy loss rate given by Eq.~\eqref{eq:eps2} to be at most the maximum allowed value given by Eq.~\eqref{eq:upper}. Since the energy loss rate is directly proportional to the $e^-e^+\to \chi\bar{\chi}$ cross section and $\sigma_{e^-e^+\to \chi\bar{\chi}} \propto \Lambda^{-4}$ [cf.~Eqs.~\eqref{eqn:totalcross-section-cooling-scalar}-\eqref{eqn:totalcross-section-cooling-tensor}], the condition $\dot{\varepsilon}\leq  10^{19}~{\rm erg}\cdot{\rm g}^{-1}\cdot{\rm s}^{-1}$ imposes a lower limit on $\Lambda$. We find that for a benchmark value of $m_\chi=30$ MeV, the undeformed ($q=1$) scenario yields a lower bound on $\Lambda$ of 2.78, 2.99 and 3.00 TeV for the S-P, V-A and T-AT type operators, respectively, while the deformed scenario with $q=1.1$ yields a stronger lower bound on $\Lambda$ of 11.91, 12.80 and 12.81 TeV, respectively. These results are summarized in Table~\ref{Table:Scalar-Vector-Tensor-30}. The supernova cooling bounds on $\Lambda$ are almost independent of the DM mass for $m_\chi\leq T_{\rm SN}$, whereas for $m_{\chi}\gg T_{\rm SN}$, the DM production in the supernova core is phase-space suppressed, leading to the exponential weakening of the bound, as is evident from the log-log plot in Fig.~\ref{fig:lambda-mchi1-T30-q1.0}.  

\begin{table}[t!]
\caption{ The lower bound on $\Lda$ from SN1987A cooling and upper bounds from free-streaming of DM 
fermions and relic density for a DM mass $m_\chi = 30$ MeV in both undeformed ($q=1$) and deformed 
($q=1.1$) scenarios in the Tsallis statistics framework. The corresponding lower bounds from the current
direct detection experiments XENON10~\cite{Essig:2012yx} and SuperCDMS~\cite{Agnese:2018col} data, as well 
as from the internal heat flux of earth~\cite{Chauhan:2016joa} are also shown, together with the future 
lower bound from LBECA~\cite{LBECA}. The lower bounds obtained from the indirect detection 
experiment Voyager1~\cite{Boudaud:2016mos} have also been listed. Note that the bound on $\Lda$ that 
follows from AMS-02 data is obtained only in the case 
of heavier DM mass i.e. $m_\chi \ge 1~\rm{TeV}$ or above~\cite{Boudaud:2016mos} and hence  is not 
shown here. The parameters used in this analysis are:  $T_{\rm SN}=30~{\rm MeV},~\rho_{\rm SN}=3\times 10^{14}~\rm{gm\: cm}^{-3}$ and $n_e=10^{37}~\rm{cm}^{-3}$. }
\label{Table:Scalar-Vector-Tensor-30}
\begin{center}
\begin{tabular}{c| c c c c c c}
\hline\hline
 $m_{\chi}=30~\rm{MeV}$ & \multicolumn{2}{c}{S-P type} & \multicolumn{2}{c}{V-A type} & \multicolumn{2}{c}{T-AT type} \\
\hline\hline

& $q=1.0$ & $q=1.1$ &  $q=1.0$ & $q=1.1$ & $q=1.0$ & $q=1.1$\\

\cline{2-7}
\\
 SN Cooling &  $2.78~ $TeV &  $11.91~ $TeV & $2.99~ $TeV & $12.8~ $TeV & $3.00~ $TeV &  $12.81~ $TeV  \\

\\ 
Free-streaming  & $0.63~ $TeV & $0.59~ $TeV & $1.12~ $TeV & $1.02~ $TeV & $1.21~ $TeV & $1.12~ $TeV \\

\\
Relic Bound & $0.072~ $TeV & $0.082~ $TeV & $0.078~ $TeV & $0.088~ $TeV & $0.082~ $TeV & $0.091~ $TeV \\ \hline

\\
XENON10  & \multicolumn{2}{c}{$ 0.004~ $TeV} & \multicolumn{2}{c}{$ 0.0061~ $TeV} & \multicolumn{2}{c}{$ 0.008~ $TeV} \\

\\
SuperCDMS  & \multicolumn{2}{c}{$ 0.0008~ $TeV} & \multicolumn{2}{c}{$ 0.0011~ $TeV} & \multicolumn{2}{c}{$ 0.0014~ $TeV} \\

\\
LBECA  & \multicolumn{2}{c}{$ 0.056~ $TeV} & \multicolumn{2}{c}{$ 0.08~ $TeV} & \multicolumn{2}{c}{$ 0.11~ $TeV} \\ \hline

\\
Voyager1  & \multicolumn{2}{c}{$ 0.192~ $TeV} & \multicolumn{2}{c}{$ 0.208~ $TeV} & \multicolumn{2}{c}{$ 0.211~ $TeV} \\

\\
Earth Heat Flux  & \multicolumn{2}{c}{$ 0.83 \times 10^{-4}~ $TeV} & \multicolumn{2}{c}{$ 1.18 \times 10^{-4}~ $TeV} & \multicolumn{2}{c}{$ 1.55 \times 10^{-4}~ $TeV} \\
\hline\hline
\end{tabular}
\end{center}
\end{table}  

\subsection{Free-streaming Bound} 
The free-streaming bounds (green lines) are obtained from the optical depth criterion 
[cf.~Section~\ref{sec:optical}], i.e. by requiring that the mean free path of the DM produced inside
the supernova should satisfy the condition given in Eq.~\eqref{Eqn:free-streaming}. Since the mean free 
path $\lambda_\chi$ is inversely proportional to the $e\chi\to e\chi$ cross section 
[cf.~Eq.~\eqref{mean_free_path}] and $\sigma_{e\chi\to e\chi}\propto \Lambda^{-4}$
[cf.~Eqs.~\eqref{eq:free1}-\eqref{eq:free3}], the condition~\eqref{Eqn:free-streaming} can be used to 
derive an upper bound on $\Lambda$ to be used in conjunction with the lower bound on $\Lambda$ from 
supernova cooling. In other words, the $\Lambda$ values above the green lines and below the red lines 
are excluded, as shown by the green-shaded regions in Fig.~\ref{fig:lambda-mchi1-T30-q1.0}. For the 
$\Lambda$ values below the green lines, the DM particles do not satisfy the 
condition~\eqref{Eqn:free-streaming}, i.e. they do not free-stream away and are always trapped inside 
the supernova core, thus invalidating the cooling bound. We find that for a benchmark value of 
$m_\chi=30$ MeV, the undeformed ($q=1$) scenario, we can rule out the $\Lambda$ values between 
0.63-2.78, 1.12-2.99 and 1.21-3.00 TeV for the S-P, V-A and T-AT type operators, respectively, 
while the deformed scenario with $q=1.1$ yields a wider exclusion region for $\Lambda$ between 
0.59-11.91, 1.02-12.8 and 1.12-12.81 TeV, respectively [see Table~\ref{Table:Scalar-Vector-Tensor-30}].
\footnote{The SN1987A cooling and free-streaming bounds on $\Lda$ in this work 
are found to be a few orders of magnitude smaller than those obtained in earlier 
works~\cite{Kadota:2014mea, Guha:2015kka} for magnetic and electric dipole moment operators. This follows 
from the fact that operators used earlier scale as $\Lda^{-1}$, while in this work, our leptophilic 
operators given by Eq.~\eqref{eq:eft} scale as $\Lda^{-2}$.}  
Just like in the cooling case, the free-streaming bounds on $\Lambda$ are almost independent of the DM mass for $m_\chi\leq T_{\rm SN}$, whereas for $m_{\chi}\gg T_{\rm SN}$, the DM production and number density in the supernova core is phase-space suppressed, leading to the exponential weakening of the bound, as is evident from the log-log plot in Fig.~\ref{fig:lambda-mchi1-T30-q1.0}. Note that the mild $q$-dependence of the free-streaming bound comes from the modified distribution function [cf.~Eq.~\eqref{eq:fq}] for the DM number density. 

\subsection{Relic Density Bound} 
The relic density bounds (blue lines) are obtained from Eq.~\eqref{Eqn:relic-density} which should match the observed DM relic density. Since $\Omega_\chi$ is directly proportional to the abundance ratio $Y_0$ which is inversely proportional to the annihilation cross section $\sigma_{\chi\bar\chi\to e^-e^+}$ [cf.Eq.~\eqref{eqn:Y0}], which in turn goes like $m_\chi^2/\Lambda^{4}$ [cf.~Eqs.~\eqref{eqn:totalcross-section-relic-scalar}-\eqref{eqn:totalcross-section-relic-tensor}], the requirement that the DM should not overclose the universe imposes an upper bound on $\Lambda$. Note that the observed DM relic density is exactly reproduced only along the blue lines for the corresponding operator type. However, the regions below the blue lines are still allowed in the sense that the missing amount of DM to explain the observed relic density could be obtained by some other means, e.g. by invoking a multi-component, freeze-in or non-thermal DM scenario. In any case, for a thermal relic DM with a benchmark $m_\chi=30$ MeV, we find that the undeformed ($q=1$) scenario yields an upper bound on $\Lambda$ of 0.072, 0.078 and 0.082 TeV for the S-P, V-A and T-AT type operators, respectively, while the deformed scenario with $q=1.1$ yields a slightly weaker upper bound on $\Lambda$ of 0.082, 0.088 and 0.091 TeV, respectively [see Table~\ref{Table:Scalar-Vector-Tensor-30}]. The bound on $\Lambda$ increases with the DM mass, because the annihilation cross section $\sigma_{e^-e^+\to \chi\bar{\chi}} \sim m_\chi^2/\Lambda^4$, and therefore, to satisfy the observed relic density for a higher DM mass, a correspondingly higher $\Lambda$ value is needed. The $q$-dependence of the relic density constraint comes from the distribution functions in the thermal averaged annihilation rate [cf.~Eq.~\eqref{eq:sigmav}], as well as the effective degrees of freedom $g_\star$ in Eq.~\eqref{eqn:Y0}. A detailed discussion of the $q$ and $T$ dependence of $g_\star$ is presented in Appendix~\ref{app:eff}. 

\subsection{Other Experimental Constraints} 
For comparison, we also translate the direct detection constraints on DM-electron coupling from electron recoil data in XENON10~\cite{Essig:2012yx}, SuperCDMS~\cite{Agnese:2018col} and SENSEI~\cite{Crisler:2018gci} onto lower limits on $\Lambda$ in Fig.~\ref{fig:lambda-mchi1-T30-q1.0} (magenta, pink and violet shaded regions, respectively). Specifically, we have used the experimental upper limits on the scattering cross section $\sigma_{e\chi\to e\chi}$ (for momentum-independent form factor) and compared them with the theoretical predictions given by Eqs.~\eqref{eq:free1}-\eqref{eq:free3} with appropriately rescaled $s\simeq (m_e+m_\chi)^2$. Similar limits can be obtained from XENON100~\cite{Aprile:2016wwo} and DarkSide-50~\cite{Agnes:2018oej} data, which are not shown here. The XENON10 bound is found to be stronger than the SuperCDMS and SENSEI bounds, but  weaker than the astrophysical bounds derived here. However, future limits, such as from LBECA (brown lines)~\cite{LBECA} could be comparable to or better than the astrophysical limits in part of the parameter space. 

Another lower limit on $\Lambda$ can be obtained from the constraints due to the heat flux of 
earth~\cite{Chauhan:2016joa}, as shown by the cyan shaded region in Fig.~\ref{fig:lambda-mchi1-T30-q1.0}. 
For this case also we have used the experimental upper limits on the scattering cross section 
$\sigma_{e\chi\to e\chi}$ and compared them with the theoretical predictions given by 
Eqs.~\eqref{eq:free1}-\eqref{eq:free3} with appropriately rescaled $s\simeq (m_e+m_\chi)^2$. 
See Table~\ref{Table:Scalar-Vector-Tensor-30} for a quantitative comparison of the direct detection 
bounds with the astrophysical ones for a benchmark $m_\chi=30$ MeV.

We have also converted the indirect detection constraints obtained from cosmic-ray measurements by AMS-02 
and Voyager1~\cite{Boudaud:2016mos} onto lower limits on $\Lambda$ and they are shown in 
Fig.~\ref{fig:lambda-mchi1-T30-q1.0} by the gray and yellow shaded regions, respectively. Here, we 
have used the experimental upper limits on the thermal average of the annihilation cross section 
times relative velocity $\langle \sigma v \rangle$ 
(with $ \sigma \equiv \sigma_{\chi \bar{\chi} \to e^+ e^-} $) and compared them with the theoretical 
predictions given by Eqs.~\eqref{eqn:totalcross-section-relic-scalar}-\eqref{eqn:totalcross-section-relic-tensor} 
and Eq. \eqref{eqn:thermal_averaged_crosssection_times_velocity}. The purple shaded region is excluded from
CMB spectral distortion data as measured by FIRAS and the pink curves are the projected CMB limits from 
PIXIE~\cite{Ali-Haimoud:2015pwa, Cappiello:2018hsu}.

There also exist collider constraints on the leptophilic DM scenario. For instance, the effective 
Lagrangian~\eqref{eq:eft} would give rise to the process $e^-e^+\to \chi\bar{\chi}$, which may leave observable signatures in the energy and momentum spectra of the objects (such as photons radiated from the electron or positron leg) recoiling against the DM. The current limits from LEP~\cite{Chae:2012bq, Fox:2011fx, Freitas:2014jla} are found to be much weaker than those shown in Fig.~\ref{fig:lambda-mchi1-T30-q1.0}, whereas the future limits from ILC~\cite{Chae:2012bq, Freitas:2014jla} could be more competitive. 

Similarly, the leptophilic operator~\eqref{eq:eft} would also induce an effective DM-quark coupling, but only at the loop-level~\cite{Kopp:2009et}. Therefore, the hadron collider constraints on monojet/monophoton, as well as the direct detection constraints from nuclear recoil are suppressed, in comparison to the bounds shown in Fig.~\ref{fig:lambda-mchi1-T30-q1.0}.

\subsection{Cooling and Free-streaming at the Crust} 
\begin{figure}[t!]
\centering
\includegraphics[width=0.49\textwidth]{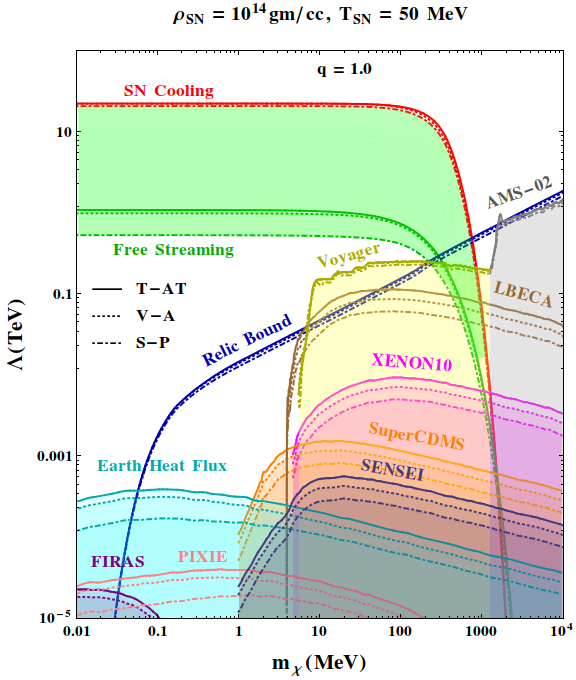}
\includegraphics[width=0.49\textwidth]{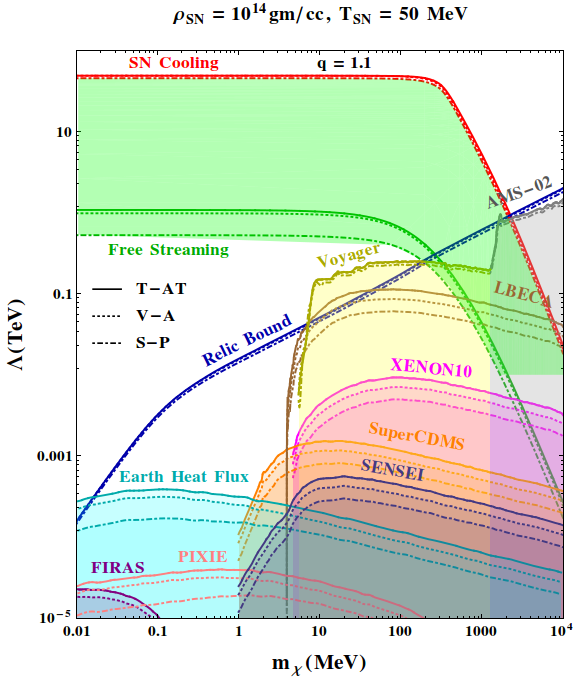}
\caption{Same as in Fig.~\ref{fig:lambda-mchi1-T30-q1.0}, but with $T_{SN} = 50~\rm{MeV}$, $\rho_{\rm SN}=10^{14}~{\rm gm\: cm}^{-3}$ and $n_e=8.7\times 10^{37}~{\rm cm}^{-3}$.}
\label{fig:lambda-mchi1-T50-q1.0}
\end{figure}

As mentioned in Section~\ref{sec:optical}, the DM fermions produced in the outer $10\%$ (i.e. crust region) of the supernova free-stream away without any hindrance, whereas the ones produced deep inside are trapped. In this section, we rederive the bounds shown in Fig.~\ref{fig:lambda-mchi1-T50-q1.0} for the case when the DM production mechanism is mostly confined to the crust, which has a higher electron density than the core. The average crust temperature is taken to be $T_{\rm SN} = 50~\rm{MeV}$, mean matter density $\rho_{\rm SN} = 10^{14}~{\rm gm\: cm}^{-3}$ and mean electron density $n_e = 8.7 \times 10^{37}~{\rm cm}^{-3}$~\cite{Raffelt:1996wa}. Our  results are shown in Fig.~\ref{fig:lambda-mchi1-T50-q1.0}. The only significant change in this case compared to Fig.~\ref{fig:lambda-mchi1-T30-q1.0} is for the supernova cooling bounds which become stronger due to the reduced mean matter density and enhanced average temperature at the crust [cf.~Eq.~\eqref{eq:eps2}]. 
A quantitative comparison of all the limits for a benchmark value of $m_\chi=30$ MeV are presented in Table~\ref{Table:Scalar-Vector-Tensor-50}.

\begin{table}[t!]
\caption{Same as in Table~\ref{Table:Scalar-Vector-Tensor-30}, but with $T_{\rm SN}=50~\rm{MeV},~ \rho_{\rm SN}= 10^{14}~\rm{gm\: cm^{-3}}$ and $n_e= 8.7\times 10^{37}~\rm{cm^{-3}}$.}
\label{Table:Scalar-Vector-Tensor-50}
\begin{center}
\begin{tabular}{c| c c c c c c}
\hline\hline
 $m_{\chi}=30~\rm{MeV}$ & \multicolumn{2}{c}{S-P type} & \multicolumn{2}{c}{V-A type} & \multicolumn{2}{c}{T-AT type} \\
\hline\hline

& $q=1.0$ & $q=1.1$ &  $q=1.0$ & $q=1.1$ & $q=1.0$ & $q=1.1$\\

\cline{2-7}
\\
 SN Cooling &  $20.31~ $TeV &  $45.35~ $TeV & $21.84~ $TeV & $48.74~ $TeV & $21.89~ $TeV &  $48.78~ $TeV  \\

\\ 
Free-streaming  & $0.49~ $TeV & $0.48~ $TeV & $0.9~ $TeV & $0.86~ $TeV & $0.98~ $TeV & $0.94~ $TeV \\

\\
Relic Bound & $0.072~ $TeV & $0.082~ $TeV & $0.078~ $TeV & $0.088~ $TeV & $0.082~ $TeV & $0.091~ $TeV \\ \hline

\\
XENON10  & \multicolumn{2}{c}{$ 0.004~ $TeV} & \multicolumn{2}{c}{$ 0.0061~ $TeV} & \multicolumn{2}{c}{$ 0.008~ $TeV} \\

\\
SuperCDMS  & \multicolumn{2}{c}{$ 0.0008~ $TeV} & \multicolumn{2}{c}{$ 0.0011~ $TeV} & \multicolumn{2}{c}{$ 0.0014~ $TeV} \\

\\
LBECA  & \multicolumn{2}{c}{$ 0.056~ $TeV} & \multicolumn{2}{c}{$ 0.08~ $TeV} & \multicolumn{2}{c}{$ 0.11~ $TeV} \\ \hline

\\
Voyager1  & \multicolumn{2}{c}{$ 0.192~ $TeV} & \multicolumn{2}{c}{$ 0.208~ $TeV} & \multicolumn{2}{c}{$ 0.211~ $TeV} \\

\\
Earth Heat Flux  & \multicolumn{2}{c}{$ 0.83 \times 10^{-4}~ $TeV} & \multicolumn{2}{c}{$ 1.18 \times 10^{-4}~ $TeV} & \multicolumn{2}{c}{$ 1.55 \times 10^{-4}~ $TeV} \\

\hline\hline
\end{tabular}

\end{center}
\end{table}

\subsection{Uncertainty on $\Lambda$ Bound}

There is some uncertainty in the mass of the progenitor of SN1987A 
 and the neutrino flux estimation~\cite{Chang:2018rso, Mahoney:2017jqk, Keil:2002in}. This 
 amounts to an uncertainty in the knowledge of the core temperature and the matter density of 
 SN1987A. Supernova simulations~\cite{Char:2015nea} show the radial variation of the 
 temperature and matter density inside the supernova, which can be encapsulated by the following simple supernova profile~\cite{Raffelt:1996wa, Chang:2016ntp}:
\bea
\rho(r) \ & = \ \rho_c \times \begin{cases} 1+ k_\rho \left( 1 - r/R_c\right);~r<R_c \\
\left(r/R_c \right)^{-\nu};~r \geq R_c 
\end{cases} , \nonumber \\
T(r) \ & = \ T_c \times \begin{cases} 1+ k_T \left( 1 - r/R_c\right);~ r<R_c \\
\left(r/R_c \right)^{-\nu/3} ;~ r \geq R_c
\end{cases} , 
\label{eq:SN-profile}
\eea 
with $R_c, ~T_c, ~\rho_c$ the core radius, core temperature and core density, respectively. For the fiducial case~\cite{Chang:2016ntp}, we have the parameter values $k_\rho =0.2, ~k_T=-0.5, ~\nu=5$ and $R_c=10~{\rm km}, ~T_c=30~\rm{MeV}, ~\rho_c=3\times 10^{14}~\rm{gm \: cm^{-3}}$. 
 
Taking into account these different environmental conditions, we have investigated how the 
bounds on the EFT scale $\Lambda$ change with respect to the case with constant core temperature and density as discussed above. We find that there is a small change  in the free-streaming bound, but the SN cooling bounds can vary up to an order of magnitude. Clearly from accurate supernova modeling, a more precise understanding of the
supernova progenitor will improve the robustness of these constraints. 

\section{Conclusion}\label{sec:conc}
We have derived model-independent astrophysical constraints on leptophilic, light, fermionic DM in the 
framework of Tsallis statistics. In particular, we have considered the DM production in supernova core 
through its dimension-six, four-Fermi interactions with electrons and positrons, and the subsequent 
constraints ensuing from supernova cooling due to energy taken away by escaping DM and free-streaming 
of DM from the outer crust. We find that for an average supernova core temperature of $T_{\rm SN}=30$ MeV, 
the SN1987A cooling imposes a lower bound on the effective cut-off scale $\Lambda \gtrsim 3$ TeV for the 
undeformed ($q=1.0$) case and 12 TeV for the deformed case with $q=1.1$ for $m_\chi<T_{\rm SN}$ 
(see Fig.~\ref{fig:lambda-mchi1-T30-q1.0} and Table~\ref{Table:Scalar-Vector-Tensor-30}). 
The corresponding limits from the free-streaming and optical depth criteria rule out the $\Lambda$ values above $\sim 0.6-1$ TeV up to the cooling bound. Restricting the DM production to the outer crust region with a higher $T_{\rm SN}=50$ MeV 
yields stronger cooling bounds on $\Lambda\gtrsim 20$ TeV (45 TeV) for $q=1.0~(1.1)$ 
(see Fig.~\ref{fig:lambda-mchi1-T50-q1.0} and Table~\ref{Table:Scalar-Vector-Tensor-50}). 
For $m_\chi \gg T_{\rm SN}$, the cooling and free-streaming bounds get significantly weaker due 
to the Boltzmann-suppression of the DM production and scattering rate in the supernova core. 
On the other hand, satisfying the observed relic density imposes an upper bound on $\Lambda$, which 
increases with DM mass, and together with the cooling/free-streaming bounds, disfavors a wide parameter range for light leptophilic DM (see Figs.~\ref{fig:lambda-mchi1-T30-q1.0} and \ref{fig:lambda-mchi1-T50-q1.0}). 
\acknowledgments
BD thanks Filippo Sala for an important comment on the free-streaming bound. PKD thanks KC Kong and Doug McKay for several useful discussions. PKD would also like to thank the Department of Physics at Washington University, 
St. Louis, for arranging a visit while this work was in progress. PKD and BD would like to thank the organizers of WHEPP2018 at IISER, Bhopal, where this work was initiated. The work of BD is supported by the US Department of Energy under Grant No. DE-SC0017987. 
The work of PKD is partly 
supported by the SERB Grant No. EMR/2016/002651. 
\section*{Appendix}
\appendix

\section{Upper limit on $q$} \label{app:q}

Here, we compare the spectrum of the emitted (anti)neutrinos from recent state-of-the-art supernova simulations (see e.g., Ref.~\cite{Tamborra:2012ac} for a time-averaged fit and Ref.~\cite{Nikrant:2017nya} for a time-dependent fit) with the Tsallis spectrum to derive an upper limit on the value of $q$. Comparing the $q$-deformed distribution function given by Eq.~\eqref{eq:fq} for fermions up to the first order with the neutrino distribution function used in Refs.~\cite{Tamborra:2012ac, Nikrant:2017nya}: 
 \begin{align}
 f_\nu \ \propto \ E^{\alpha} \exp{\left[-(\alpha + 1)\frac{E}{E_{\rm av}}\right]} \, ,
 \end{align}
where $E_{\rm av}$ is the average neutrino energy, and $\alpha$ is the neutrino-spectrum shape parameter (with $2 \le \alpha \le 4$ providing a good fit to the spectrum),  we get a relation between $q$ and $\alpha$: 
\bea
\frac{1}{4-3q} &\ = \ & \alpha +1, \qquad {\rm or}, \qquad q \ = \ \frac{4 \alpha +3}{3 \left(\alpha +1\right)} \, .
\label{eq:alphaq}
\eea 

Given the maximum allowed value of $\alpha_{\rm max}=4$ from the fits to the neutrino spectra~\cite{Tamborra:2012ac, Nikrant:2017nya}, we obtain from Eq.~\eqref{eq:alphaq} the corresponding value for $q_{\rm max}=1.27$, which can be considered as the limiting value of the deformation parameter $q$.  Hence, the benchmark value of $q=1.1$ used in our present analysis for the $q$-deformed scenario is consistent with the state-of-the-art supernovae modeling.  

\section{Effective Degrees of Freedom } \label{app:eff}
To better understand the relic density bound on $\Lda$ as a function of $m_\chi$ and its $q$-dependence, we briefly discuss the temperature and $q$-dependence of the effective degrees of freedom $g_*$, starting from early times before the electroweak phase transition to the 
present day. We take the relevant temperature range to be T=(1 keV, 10 TeV), which encapsulates all the essential features we want to study here. From the excellent agreement of the cosmic microwave background with a black body spectrum~\cite{Aghanim:2018eyx}, we know that the early universe was close to thermal equilibrium. Using statistical mechanics, we can 
calculate the energy density, pressure and entropy density for a system in equilibrium at any given temperature by simply summing up the contributions from all the particle species present in the thermal bath. The contribution from a certain particle species depends on its mass and intrinsic degrees of freedom (such as spin and isospin degeneracy). All this information can be incorporated via the temperature-dependent effective degree of freedom $g_*(T)$, defined relative to the photon. We have four different $g_*$ functions, i.e., $g_{*n}$, $g_{*\epsilon}$, $g_{*p}$ and $g_{*s}$ related to the number density, energy density, pressure and entropy density, 
respectively. By including the intrinsic degrees of freedom ($g_j$) for each particle species $j$, the total effective degrees of freedom for the four kinds mentioned above are given by~\cite{Husdal:2016haj, Ryden} 
 \begin{align}
g_{*n} \ & = \ \sum\limits_j\frac{g_j}{2\zeta(3)}\int\limits_{z_j}^\infty \frac{u\sqrt{u^2-z_j^2}}{e^u\pm 1}du \, , \label{gstarn}\\
 g_{*\epsilon} \ & =  \ \sum\limits_j\frac{15g_j}{\pi^4} \int\limits_{z_j}^{\infty} \frac{u^2 \sqrt{u^2-{z_j}^2}}{e^u \pm 1} du \, ,  \label{gstare} \\
  g_{*p} \ & = \  \sum\limits_j \frac{15g_j}{\pi^4} \int\limits_{z_j}^{\infty} \frac{(u^2-{z_j}^2)^{3/2}}{e^u \pm 1} du \, ,  \label{gstarp} \\
 g_{*s} \ & = \ \frac{ 3 g_{*\epsilon} +  g_{*p}}{4} \, , \label{gstars}
 \end{align}
where $z_j=m_j/kT$, $\zeta(3)$ is the Riemann zeta function of argument 3, and the $+~(-)$ sign is for fermions (bosons). 
The quantity $g_*$ that appears in Eq.~\eqref{eqn:Y0} for the relic density calculation is only related to the entropy and energy densities~\cite{Gondolo:1990dk}:  
  \bea
  g_*^{1/2} (T) \ = \ \frac{g_{*s}}{g^{1/2}_{*\epsilon}} \left( 1 + \frac{1}{3} \frac{T}{g_{*s}} 
  \frac{dg_{*s}}{dT} \right) \, .
  \label{gstarhalf}
  \eea
In the $q$-deformed Tsallis statistics formalism, we just replace the normal exponential function $e^u$ in Eqs.~\eqref{gstarn}-\eqref{gstars} by the $q$-deformed exponential $e^u_q$ [cf.~Eq.~\eqref{eq:piq}] and 
accordingly the expressions for $g_{*s}$ and $g_{*\epsilon}$ get modified.  Thus, the temperature-dependent, $q$-deformed effective degree of freedom entering Eq.~\eqref{eqn:Y0} is given by  
  \bea
  g^{1/2}_{*}(q,T) \ = \ \frac{g_{*s,q}}{g^{1/2}_{*\epsilon,q}} \left( 1 + \frac{1}{3} \frac{T}{g_{*s,q}} 
  \frac{dg_{*s,q}}{dT} \right) \, .
  \label{gstarhalfq}
  \eea
Note that in the limit $q \to 1$, we have $e^u_q \to e^u$ and we recover the undeformed 
effective degrees of freedom defined in Eq.~\eqref{gstarhalf}. 
\begin{figure}[t!]
\centering
\includegraphics[width=0.49\textwidth]{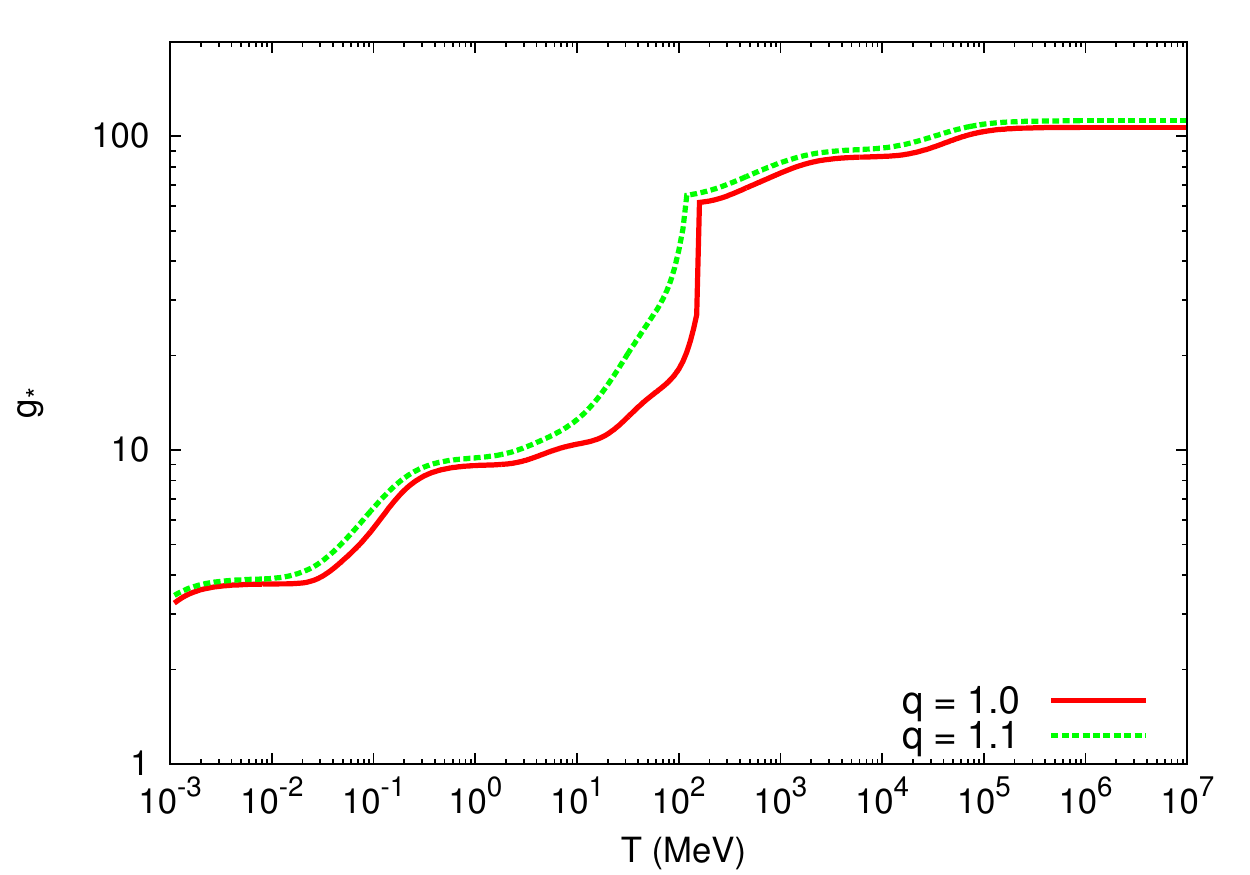}
\includegraphics[width=0.49\textwidth]{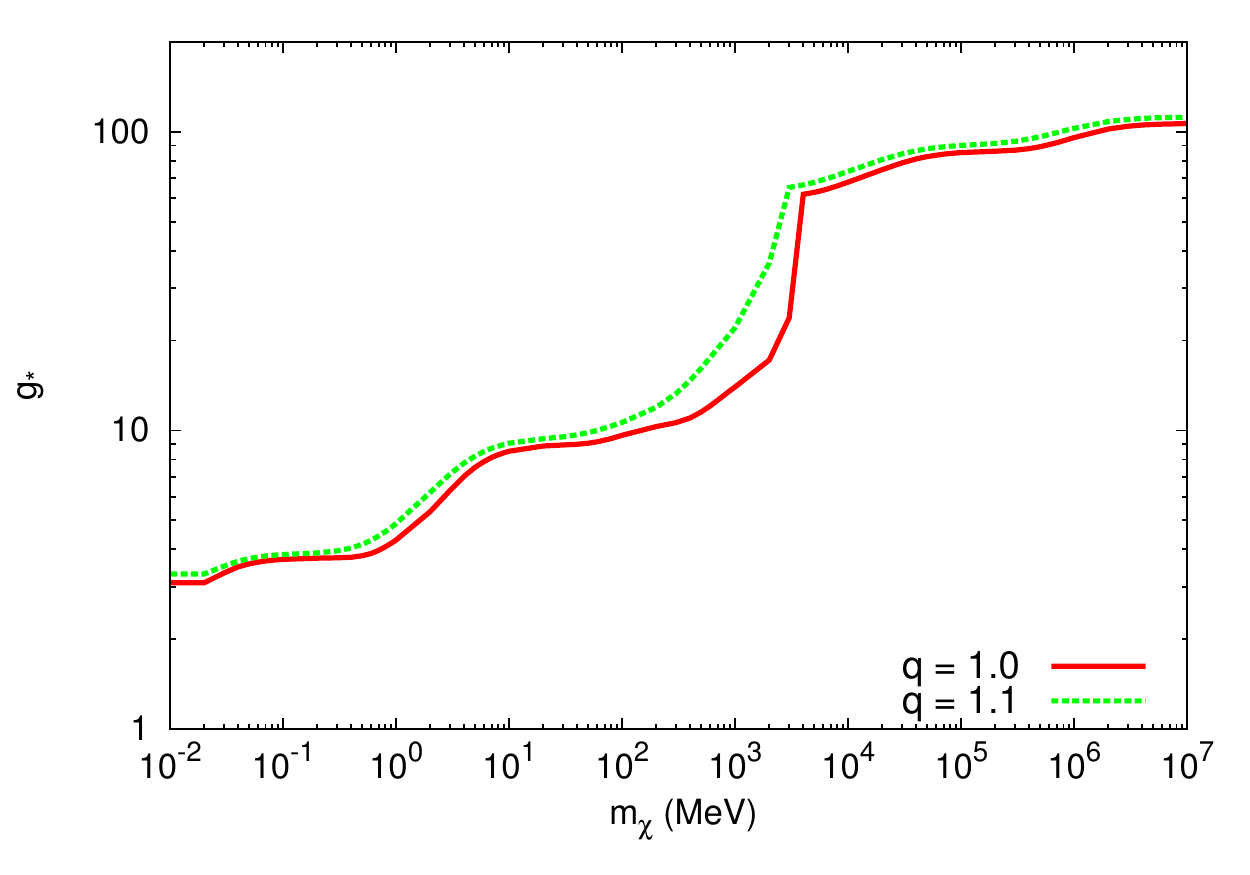}
\caption{Variation of the effective degrees of freedom $g_*$ with temperature $T$ (left panel) for a DM mass $m_\chi=20 \: T$  and with DM mass $m_\chi$ (right panel) for fixed $T=T_f$ (freeze-out temperature) for deformed ($q=1.1$) and undeformed ($q=1.0$) scenarios.}
\label{fig:gstarvs-temp}
\end{figure}

In Fig.~\ref{fig:gstarvs-temp} left panel, we have shown the variation of $g_*$ with $T$ in our leptophilic DM setup with a DM mass $m_\chi= 20\: T$ in order to ensure that the DM has frozen out for the temperature of interest and the only degrees of freedom contributing to $g_*$ are the SM species. We have shown the results for both undeformed ($q=1.0$) and deformed ($q=1.1$) cases for comparison. The jumps in the value of $g_*$ are due to different SM species going out-of-equilibrium at that temperature. However, we find that the total effective degrees of freedom is slightly higher in the $q$-deformed case, as compared to the undeformed case, at any given temperature. Similar behavior can be seen in the right panel of Fig.~\ref{fig:gstarvs-temp}, where we plot the variation of $g_*$ with $m_\chi$ for a fixed temperature $T=T_f\simeq m_\chi/20$ (freeze-out temperature), again to ensure that the only degrees of freedom contributing to $g_*$ are the SM species. 

\begin{figure}[t!]
\centering
\includegraphics[width=0.5\textwidth]{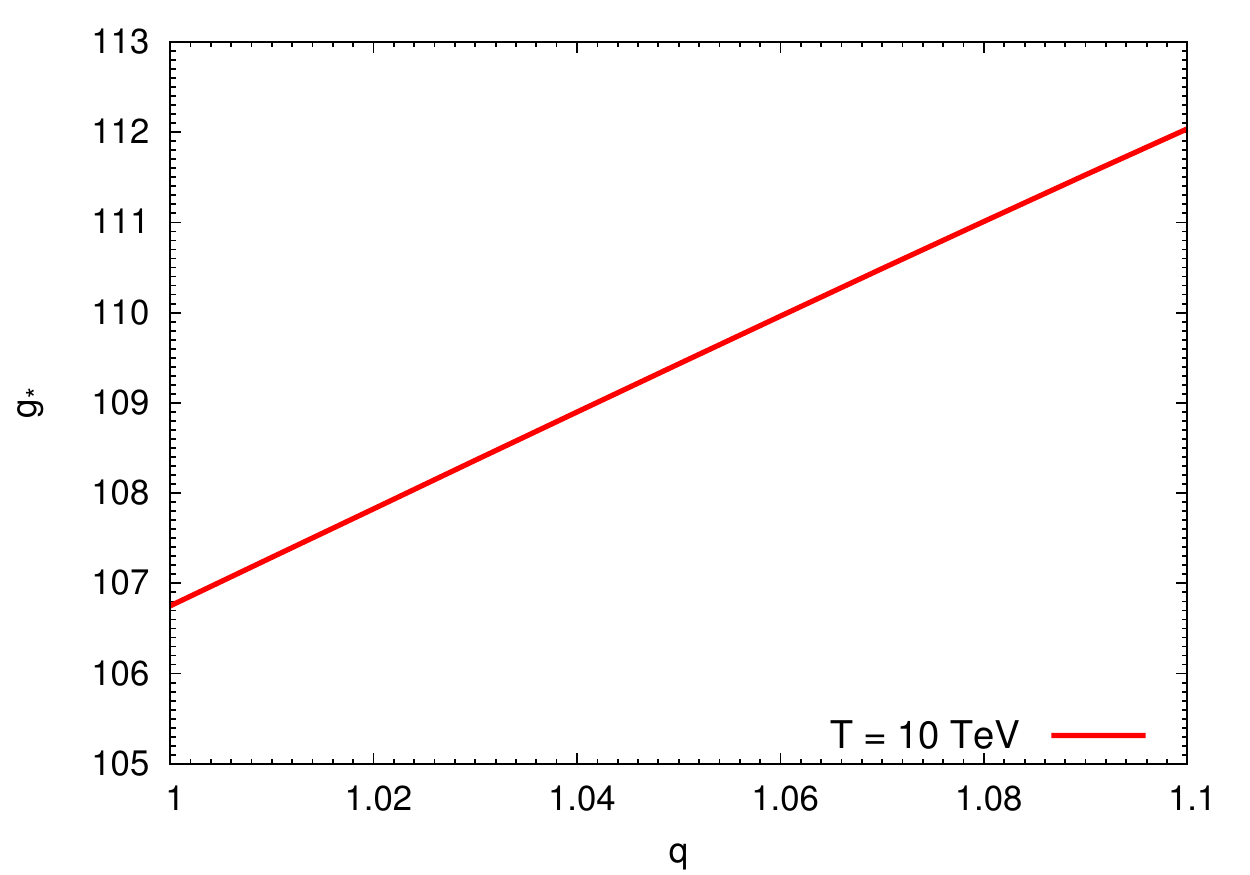}
\caption{$g_*$ variation with $q$ for $T = 10$ TeV. }
\label{fig:gstarvs-q}
\end{figure}

The variation of $g_*$ with $q$ is shown in Fig.~\ref{fig:gstarvs-q} for $T = 10$ TeV. As $q$ varies from $q=1.0$ to $q=1.1$, $g_*$ steadily increases  from $\sim 106.7$ to $112$. A change in the value 
of $T$ from 10 to 1 TeV does not lead to any significant change in the $g_*$ value, as shown in Table~\ref{table:gstartvsq-temp}  and  Fig.~\ref{fig:gstarvs-temp}). The fact that $g_*$ increases with $q$ at a given temperature $T$ is a typical feature of the underlying Tsallis statistics and is not to be interpreted as the appearance of any new particle species. 

\begin{table}[t!]
\caption{Numerical values of $g_*$ for different temperatures and $q$ values. }
\label{table:gstartvsq-temp}
\begin{center}
\begin{tabular}{c c c c c c c}
\hline\hline
 $T$  & \multicolumn{2}{c}{$q=1.0$} & \multicolumn{2}{c}{$q=1.05$} & \multicolumn{2}{c}{$q=1.1$} \\
\hline\hline
\\
10 TeV  & \multicolumn{2}{c}{$ 106.75$} & \multicolumn{2}{c}{$ 109.43$} & \multicolumn{2}{c}{$ 112.04$} \\
\\
1 TeV  & \multicolumn{2}{c}{$ 106.72$} & \multicolumn{2}{c}{$ 109.40$} & \multicolumn{2}{c}{$ 112.01$} \\
\\
%
%
%
%
\hline
\end{tabular}
\end{center}
\end{table} 

 \begin{figure}[t!]
 \centering
 \includegraphics[width=0.5\textwidth]{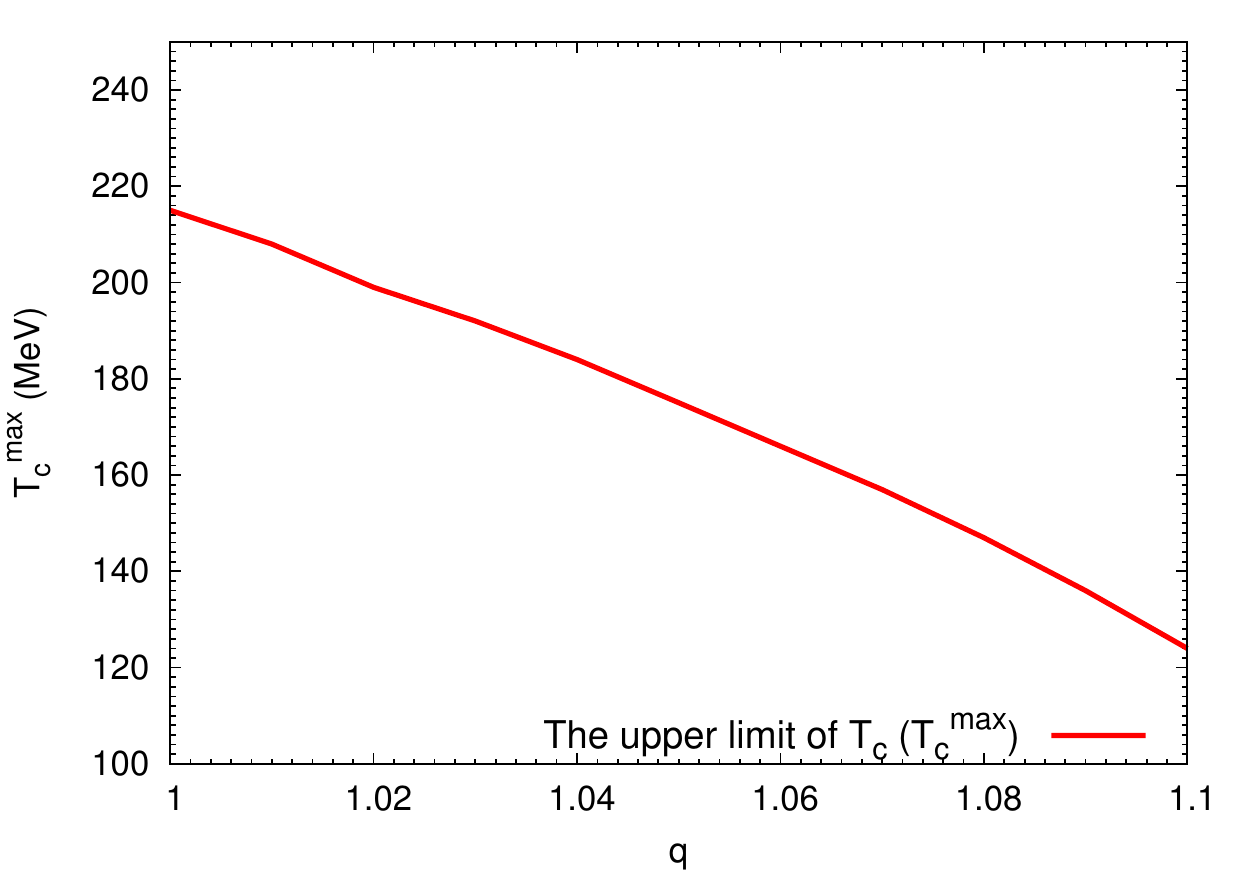}
 \caption{Upper limit on the QCD phase transition temperature $T_c^{\rm max}$ as a function of $q$.}
 \label{fig:TransTempvs-q}
 \end{figure}
One should note that the QCD phase transition temperature $T_c$ plays an important role in determining the behavior of $g_*$ during the quark-gluon plasma to hot hadron
gas transition epoch. In particular, there exists a maximum allowed value, $T_{c}^{\rm max}$, beyond which  we get a very steep, unphysical 
increase in $g_*$  due to the appearance of many heavier hadrons,
whose numbers grow almost exponentially as the temperature increases~\cite{Husdal:2016haj}. The 
exact value of this maximum phase transition temperature is found to be $q$-dependent in the 
Tsallis statistics formalism. A variation of $T_{c}^{\rm max}$ 
against $q$ is shown in Fig.~\ref{fig:TransTempvs-q}. As $q$ changes from 
$q=1.0$ to $q=1.1$, $T_{c}^{\rm max}$ is found to vary from $215$ MeV to $124$ MeV, using which we 
obtain an empirical formula 
\bea 
T_{c}^{\rm max} \ = \ 351-0.872 \: e^{5q} \, . 
\eea
Note that this is just an upper limit and the actual temperature at which the QCD phase transition occurs is expected to be lower, since there most likely are more possible hadronic states. A more accurate estimate for the transition
temperature can be obtained from  lattice Monte-Carlo simulations, which suggest $T_c=150-170$  MeV~\cite{Petreczky:2012rq} in the undeformed scenario, consistent with the upper bound derived here (see Fig.~\ref{fig:TransTempvs-q}).
\section{Analytic Expressions for Cross Sections}\label{app:cross} 
  \subsection{$e^+e^-\to \chi\bar{\chi}$}\label{app:cross1}
  \begin{align}
   \sigma_{e^-e^+ \to \chi \bar{\chi}}^{\rm S-P} \ & = \ \frac{A}{4 \pi \Lambda^4  s} \Biggl[\left(\frac{1}{2} \left\lbrace(c_S^\chi)^2+(c_P^\chi)^2\right\rbrace \left(s-2 m_{\chi}^2\right)+m_{\chi}^2 \left\lbrace(c_P^\chi)^2-(c_S^\chi)^2\right\rbrace\right) \nonumber \\ 
  & \qquad \qquad \times  \left(\frac{1}{2} \left\lbrace(c_S^e)^2+(c_P^e)^2\right\rbrace \left(s-2 m_{e}^2\right)+m_{e}^2 \left\lbrace(c_P^e)^2-(c_S^e)^2 \right\rbrace \right) \Biggr] \, , \label{eq:A1}\\
  \sigma_{e^-e^+ \to \chi \bar{\chi}}^{\rm V-A} \ & = \ \frac{A}{12 \pi \Lambda^4  s } \Biggl[ (c_V^{\chi})^2 \left(s+2 m_{\chi}^2\right) \left\lbrace(c_V^{e})^2 \left(s+2 m_{e}^2\right)+(c_A^{e})^2 \left(s-4 m_{e}^2\right) \right\rbrace  \nonumber \\  
& \qquad + (c_A^{\chi})^2 \left(s-4 m_{\chi}^2\right) \left\lbrace (c_V^{e})^2 \left(s + 2 m_{e}^2\right) + (c_A^{e})^2 \left(s - 4 m_{e}^2 \cdot \frac{s - 7 m_{\chi}^2}{s-4 m_{\chi}^2} \right)\right\rbrace \Biggr] \, ,\\
  \sigma_{e^-e^+ \to \chi \bar{\chi}}^{\rm T-AT} \ & = \ \frac{A}{6 \pi \Lambda ^4  s } \Biggl[(c_{T}^{\chi})^2 \left\lbrace  s \left(s + 2 m_{\chi}^2 \right) + 2 m_{e}^2 \left(s + 20 m_{\chi}^2\right)\right\rbrace  \nonumber \\ 
& \qquad \qquad \qquad + (c_{AT}^{\chi})^2 \left\lbrace s \left(s+2 m_{\chi}^2\right)+2 m_{e}^2 \left(s - 16 m_{\chi}^2\right) \right\rbrace \Biggr] \, , \label{eq:A3}
  \end{align}
  where $A = \sqrt{(s-4 m_{\chi}^2)/(s-4 m_{e}^2)}$ and $s$ is the center-of-mass energy. 
  
\subsection{$e^-\chi\to e^-\chi$} \label{app:cross2}
\begin{align}
\sigma_{e \chi \to e  \chi }^{\rm S-P} \ & = \  \frac{1 }{48 \pi \Lambda^4  s^3 } \Biggl[ (c_S^\chi)^2 \biggl\lbrace \left( s^4 + 2 s^3 m_\chi^2 -6 s^2 m_\chi^4  + 2 s m_\chi^6 + m_\chi^8 \right) \left\{(c_S^{e})^2  + (c_P^{e})^2 \right\} \nonumber \\ 
& \qquad \qquad \qquad + 3 s m_e^2 \left( s^2 + 6 s m_\chi^2 + m_\chi^4 \right) \left\{(c_S^{e})^2  - (c_P^{e})^2 \right\} \biggr\rbrace 
 \nonumber \\ 
   \displaybreak & \qquad + (c_P^\chi)^2 (s - m_\chi^2)^2 \biggl\lbrace \left(s - m_\chi^2 \right)^2 \left\lbrace(c_S^{e})^2  + (c_P^{e})^2 \right\rbrace + 3 s m_e^2 \left\lbrace(c_S^{e})^2  - (c_P^{e})^2 \right\rbrace \biggr\rbrace  \Biggr]\, , \label{eq:A4} \\
\sigma_{e \chi \to e  \chi }^{\rm V-A} \ & = \ \frac{1 }{24 \pi \Lambda^4  s^3 } \Biggl[ (c_V^\chi)^2 \biggl\lbrace \left(4 s^4 - 10 s^3 m_\chi^2 +9 s^2 m_\chi^4  -4 s m_\chi^6 + m_\chi^8 \right) \left\lbrace(c_V^{e})^2  + (c_A^{e})^2 \right\rbrace \nonumber \\ 
& \qquad \qquad \qquad \qquad \qquad - 3 s m_e^2 \left( s^2 - 6 s m_\chi^2 + m_\chi^4 \right) \left\lbrace(c_V^{e})^2  - (c_A^{e})^2 \right\rbrace \biggr\rbrace 
\nonumber \\ 
& \qquad \qquad + (c_A^\chi)^2 \biggl\lbrace \left(4 s^4 - 4 s^3 m_\chi^2 -3 s^2 m_\chi^4  +2 s m_\chi^6 + m_\chi^8 \right) \left\lbrace(c_V^{e})^2  + (c_A^{e})^2 \right\rbrace \nonumber \\ 
& \qquad \qquad \qquad \qquad \qquad- 3 s m_e^2 \left( s^2 +10 s m_\chi^2 + m_\chi^4 \right) \left\lbrace(c_V^{e})^2  - (c_A^{e})^2 \right\rbrace \biggr\rbrace  \Biggr] \, , \\
\sigma_{e \chi \to e  \chi }^{\rm T-AT} \ & = \ \frac{1 }{6 \pi \Lambda^4  s^3 }  \Biggl[\left\lbrace(c_T^{\chi})^2  + (c_{AT}^{\chi})^2 \right\rbrace\left( 7 s^4 - 13s^3 m_{\chi}^2  +6 s^2m_{\chi}^4 - s m_{\chi}^6 +m_{\chi}^8 \right) \nonumber \\ & \qquad \qquad \qquad \qquad \qquad + \left\lbrace(c_T^{\chi})^2  - (c_{AT}^{\chi})^2 \right\rbrace \left( 36 s^2 m_e^2 m_\chi^2 \right) \Biggr]  \, . \label{eq:A6}
\end{align}
\subsection{$\chi\bar{\chi}\to e^+e^-$} \label{app:cross3}
\begin{align}
\sigma_{\chi \bar{\chi} \to e^-e^+}^{\rm S-P} \ & = \ \frac{1}{4 A \pi \Lambda^4  s} \Biggl[\left(\frac{1}{2} \left\{(c_S^\chi)^2+(c_P^\chi)^2\right\} \left(s-2 m_{\chi}^2\right)+m_{\chi}^2 \left\{(c_P^\chi)^2-(c_S^\chi)^2\right\}\right) \nonumber \\ 
& \qquad \qquad \times \left(\frac{1}{2} \left\{(c_S^e)^2+(c_P^e)^2\right\} \left(s-2 m_{e}^2\right)+m_{e}^2 \left\{(c_P^e)^2-(c_S^e)^2 \right\} \right) \Biggr] \, , \label{eq:A7}\\
\sigma_{\chi \bar{\chi} \to e^-e^+}^{\rm V-A}\ & = \ \frac{1}{12 A \pi \Lambda^4  s } \Biggl[ (c_V^{\chi})^2 \left(s+2 m_{\chi}^2\right) \biggl\lbrace (c_A^{e})^2 \left(s-4 m_{e}^2\right)+(c_V^{e})^2 \left(s+2 m_{e}^2\right) \biggr\rbrace  \nonumber \\  
&  \qquad 
+ (c_A^{\chi})^2 \left(s-4 m_{\chi}^2\right) \left\lbrace (c_A^{e})^2 \left(s - 4 m_{e}^2 \cdot \frac{s - 7 m_{\chi}^2}{s-4 m_{\chi}^2} \right)+ (c_V^{e})^2 \left(s + 2 m_{e}^2\right) \right\rbrace \Biggr] \\
\sigma_{\chi \bar{\chi} \to e^-e^+}^{\rm T-AT} \ & = \ \frac{1}{6 A \pi \Lambda ^4  s } \Biggl[(c_{T}^{\chi})^2 \biggl\lbrace  s \left(s + 2 m_{\chi}^2 \right) + 2 m_{e}^2 \left(s + 20 m_{\chi}^2\right)\biggr\rbrace \nonumber \\ 
& \qquad \qquad \qquad \qquad + (c_{AT}^{\chi})^2 \biggl\lbrace s \left(s+2 m_{\chi}^2\right)+2 m_{e}^2 \left(s - 16 m_{\chi}^2\right) \biggr\rbrace \Biggr] \, . \label{eq:A9}
\end{align}


\end{document}